\documentclass[]{spie}  %>>> use for US letter paper
\usepackage{amsmath,amsfonts,amssymb}
\usepackage{graphicx}
\usepackage[colorlinks=true, allcolors=blue]{hyperref}

\newcommand*\aap{A\&A}

\newcommand*\aj{AJ}

\newcommand*\apj{ApJ}

\newcommand*\araa{ARA\&A}

\newcommand*\mnras{MNRAS}

\newcommand*\nat{Nature}

\newcommand*\pasp{PASP}

\newcommand*\procspie{Proc SPIE}

\title{Exoplanet science with a space-based mid-infrared nulling interferometer}

\author[a]{Sascha P. Quanz}
\author[b]{Jens Kammerer}
\author[c]{Denis Defr\`ere}
\author[c]{Olivier Absil}
\author[a]{Adrian M. Glauser}
\author[d]{Daniel Kitzmann}

\affil[a]{ETH Zurich, Institute for Particle Physics and Astrophysics, Wolfgang Pauli Strasse 27, CH-8093 Zurich, Switzerland}
\affil[b]{Research School of Astronomy \& Astrophysics, Australian National University, Canberra, ACT 2611, Australia}
\affil[c]{Space sciences, Technologies \& Astrophysics Research (STAR) Institute, Universit\'e de Li\`ege, 19c all\'ee du Six Aout, b\^at B5c, B-4000 Li\`ege, Belgium}
\affil[d]{University of Bern, Center for Space and Habitability (CSH), Gesellschaftsstrasse 6 (G6), CH-3012 Bern, Switzerland}

\authorinfo{Further author information (e-mail): sascha.quanz@phys.ethz.ch}

\begin{document} 
\maketitle

\begin{abstract}
One of the long-term goals of exoplanet science is the (atmospheric) characterization of a large sample ($>$100) of terrestrial planets to assess their potential habitability and overall diversity. Hence, it is crucial to quantitatively evaluate and compare the scientific return of various mission concepts. Here we discuss the exoplanet yield of a space-based mid-infrared (MIR) nulling interferometer. We use Monte-Carlo simulations, based on the observed planet population statistics from the Kepler mission, to quantify the number and properties of detectable exoplanets (incl. potentially habitable planets) and we compare the results to those for a large aperture optical/NIR space telescope. We investigate how changes in the underlying technical assumptions (sensitivity and spatial resolution) impact the results and discuss scientific aspects that influence the choice for the wavelength coverage and spectral resolution. Finally, we discuss the advantages of detecting exoplanets at MIR wavelengths, summarize the current status of some key technologies, and describe what is needed in terms of further technology development to pave the road for a space-based MIR nulling interferometer for exoplanet science. 
\end{abstract}


\section{INTRODUCTION}
\label{sec:intro}
Since the first detection of a planet orbiting a main sequence star other than our Sun in 1995\cite{mayorqueloz1995} the field of exoplanet science has been growing at a breathtaking speed: to date we know more than 5000 exoplanets and exoplanet candidates\footnote{see, \url{exoplanet.org}}. The overwhelming majority of these objects were detected via dedicated long-term surveys using indirect techniques -- the radial velocity (RV) or the transit technique -- from both the ground (e.g., the HARPS survey or the California Planet Survey) and from space (e.g., NASA's Kepler mission). Thanks to the statistics derived from these surveys, we have a first quantitative understanding of the occurrence rate of different planet types as a function of their radius / mass, orbital period and also spectral type of the host star\cite{burke2015,mayor2011,bonfils2013}. %In addition, we now have first estimates for the occurrence rate of terrestrial planets lying within the classic habitable zone (HZ) of their host stars\footnote{see close-out report from NASA's ExoPAG SAG13 (\url{https://exoplanets.nasa.gov/exep/exopag/sag/})}. 
For some planets, transit spectroscopy and/or secondary eclipse spectroscopy measurements (primarily done from space with the Hubble Space Telescope and the Spitzer Space Telescope) provide empirical constraints on the atmospheric composition of these objects\cite{seager2010,sing2016}. With a few exceptions\cite{kreidberg2014,dewit2016}, up to now these investigations targeted so-called hot Jupiters, gas-giant planets on orbits with periods of a few days only, but with the upcoming launch of the James Webb Space Telescope (JWST) in 2020 some smaller planets, i.e., the ‘low-hanging fruits’ with sizes possibly down to those of super-Earths or Earth-like planets, might come within reach of atmospheric detection or even characterization studies\cite{morley2017}.

The tremendous growth in exoplanet science in the last 20 years has been driven by continuous progress in relevant technologies, but also in data calibration and analysis strategies. In particular the Kepler mission has impressively demonstrated what can be achieved with dedicated exoplanet missions if the technical requirements are stringently and without compromise derived from a well-defined mission objective.

The future for exoplanet science looks bright: on both sides of the Atlantic both NASA and ESA are preparing to launch a suite of dedicated exoplanet missions in the coming 10-15 years. Recently, NASA launched TESS (Transiting Exoplanet Survey Satellite) \cite{sullivan2015} which will search almost the whole sky for new Earth-sized and super-Earth-sized exoplanets with orbital periods $<$40-50 days around stars significantly brighter (and hence closer to Earth) than the Kepler targets. CHEOPS (Characterizing Exoplanet Satellite) \cite{cessa2017} -- the first ESA S-Class (small) Science Mission with the goal of measuring the size of known transiting planets with high accuracy and searching for transit signals of well-selected exoplanets initially discovered with the RV technique -- is foreseen to be launch-ready at the very end of 2018. In the mid 2020s, PLATO (Planetary Transits and Oscillations of stars) \cite{rauer2014} will follow as the third M-class mission in ESA’s Cosmic Vision Program. Similar to Kepler, albeit targeting brighter stars with higher precision and longer time baseline, PLATO will uncover hundreds of new Earth-sized exoplanets and provide unprecedented constraints on the occurrence rate of terrestrial planets in the habitable zone of Solar-type stars. Most recently, ESA announced that with ARIEL (Atmospheric Remote sensing Infrared Exoplanet Large survey mission) \cite{tinetti2016} another M-class exoplanet mission will follow in 2028. In addition to these missions employing the transit technique, NASA’s WFIRST-AFTA \cite{noecker2016}, if implemented, should test new coronagraphic technologies for space-based applications and reveal a few exoplanets around the nearest stars in reflected light via high-contrast direct imaging in the 2020s. Similarly, the next generation of 30-40 m ground-based extremely large telescopes (ELTs) will feature adaptive-optics assisted high-contrast imaging and high-dispersion spectroscopy instruments (such as ESO’s ELT/METIS\cite{brandl2016}, ELT/HIRES\cite{marconi2016} and ELT/PCS\cite{verniaud2010}) to directly image and/or characterize exoplanets within a few dozens parsec from the Sun\cite{quanz2015}. On the RV side, ESO is commissioning the new ESPRESSO instrument at the VLT on Cerro Paranal, which promises to achieve the required precision to detect Earth-twins around nearby bright and quiet Solar-type stars\cite{gonzalezhernandez2017}.
While the detection and characterization of exoplanets is partially defining the road-maps of major space agencies and observatories around the world, none of the missions and projects mentioned above will be able to address one of the main long-term science goals of exoplanet research: to directly characterize the atmospheric properties of a statistically significant sample ($>$100 objects) of terrestrial exoplanets in order to identify possible biosignatures and investigate the diversity of atmospheric properties in comparison with the Solar System planets and models for planet formation and evolution.

We believe that to achieve this goal a new large space-based mission is required, avoiding the perturbing effects (turbulence and thermal emission) of the Earth’s atmosphere. Given the intrinsic faintness of exoplanet signals, it seems that in-depth characterization is only feasible for objects in the immediate vicinity of the Sun ($<$20 pc), where the number of transiting planets is too small. It is hence very likely that the optimum configuration will be an imaging mission in order to observe a sizable sample of planets. However, a key question is whether to build a large, single aperture UV/optical/NIR telescope to characterize planets in reflected light or a large mid-infrared nulling interferometer to probe the planets’ thermal emission. In the context of the decadal survey NASA has ongoing studies to investigate the HabEX\cite{messesson2016} and LUVOIR\cite{peterson2017} concepts for reflected light, but -- at least to our knowledge -- there are no coordinated studies underway that address the science potential and the technical challenges of a space-based mid-infrared (MIR) nulling interferometer. In this paper, we quantify the potential scientific yield of such an MIR interferometry mission and compare it to that of a large single dish optical/NIR telescope. We further discuss the advantages of investigating the thermal emission of (small / terrestrial) exoplanets and describe potential next steps in terms of technology development. 

\section{SIMULATION APPROACH AND ASSUMPTIONS}
\label{sec:sims}
The core of our analyses are Monte Carlo simulations that are described in detail in Kammerer \& Quanz (2018)\cite{kammerer2018}. In short, for a sample of 318 stars (54 F stars, 72 G stars, 71 K stars, and 121 M stars) within 20 pc from the Sun we simulate planetary systems (5000 times for each star) in accordance with occurrence rates determined by NASA's Kepler mission and quantify how many of these planets would be detectable with a space-based MIR interferometer. We randomly draw planets from corresponding period and radius distributions and put them on random positions along their orbits. The orbits are assumed to be circular (see Kammerer \& Quanz (2018)\cite{kammerer2018} for a discussion of the impact of eccentric orbits) and the orbital plane has a random inclination. We then compute the planets equilibrium temperature and their thermal flux (assuming black-body emission and random values for the planets' Bond albedo drawn from the range of albedos observed in the Solar System), and we also determine their brightness in reflected light (assuming random values for the geometric albedo again drawn from the range observed in the Solar System). In contrast to Kammerer \& Quanz (2018)\cite{kammerer2018} we consider not only one but two different planetary occurrence rate estimates, i.e., underlying planet populations. The first population (we will refer to it as ``Population 1'') combines results from various papers and for different host star spectral types and is described in Kammerer \& Quanz (2018)\cite{kammerer2018}. The second population (referred to as ``Population 2''), recently published in Hsu et al. (2018)\cite{Hsu2018}, is based on a more comprehensive dataset from the Kepler mission (covering Q1-Q16) and focuses on solar type stars only. In this case we complement the results with those for M-stars from the first population to ensure a spectral type dependence of the planet occurrence rates. We carried out Monte Carlo simulations using both planet populations to test the impact of the underlying statistics on the derived planet detection rates.

On the technical side we assume the instrument specifications and performance as initially foreseen for the Darwin mission\cite{cockell2009} and as summarized in Kammerer \& Quanz (2018)\cite{kammerer2018}. The inner working angle (IWA) is assumed to be 5.0 mas at 10\,$\mu$m wavelength and the 10$\sigma$ sensitivity limits -- for 35'000 s integration time -- are 0.16, 0.54 and 1.39 $\mu$Jy at 5.6, 10.0 and 15.0 $\mu$m wavelength, respectively (in the baseline scenario). We do not make any explicit assumption about null-depth or contrast and ignore stellar leakage (which is primarily an issue for the nearest stars). While the sensitivities above implicitly include background noise from local zodiacal light\cite{glasse2015}, we do not yet include any additional noise terms from exozodiacal dust belts\footnote{The impact of exozodiacal dust on the exoplanet yield has been specifically addressed in Defr\`ere et al. 2010\cite{defrere2010} and Defr\`ere et al. 2012\cite{defrere2012}.}. However, we consider the used sensitivity limits to be justifiable as, for the time being, we did not carry out any overall or target-specific optimization. In addition, as we assume that the mission is split into a ``detection phase'' and a ``characterization phase'', each taking 2-3 years, we believe to have sufficient margin in the detection phase so that technical overheads and longer on-source times can be accommodated for. 

The technical assumptions for a large space-based single dish optical/NIR telescope are also given in Kammerer \& Quanz (2018)\cite{kammerer2018}: in the baseline scenario an aperture of $D$=12 m is assumed, the IWA is 2$\lambda$/$D$, the achievable contrast is kept fixed at 1e-10 irrespective of the separation and which filter is used, and the sensitivity limits are 3.31e-10, 9.12e-10, and 8.32e-10 Jy, at 0.55, 1.2 and 1.6 $\mu$m, respectively. These limits are based on the same amount of on-source time and refer to the same significance as the ones for the interferometer mentioned above. It is worth mentioning that these limits were estimated based on classical imaging applications and broadband filters; potential throughput losses due to the use of coronagraphs are not taken into account here. 

In our statistical analysis we consider a planet ``detected'' when its brightness (thermal emission or reflected light) exceeds the sensitivity limits, its separation is larger than the IWA of the respective instrument, and -- in case of the optical/NIR telescope -- its contrast is smaller than the assumed contrast limit. In the sub-sequent analyses we focus on planets with radii smaller or equal to 6 R$_{Earth}$. 

\section{RESULTS FOR DETECTION PHASE}
In this section we summarize our results for what we refer to as the ``detection phase'' of a mission. This is the first half or so of the total mission lifetime, aiming at compiling a large sample of detected exoplanets. A fraction of this sample is then followed-up in detail in the ``characterization phase'' of the mission.

\label{sec:results_det}
\subsection{Total exoplanet yield - baseline cases}
\begin{figure}[t!]
   \begin{center}
   \begin{tabular}{c} %% tabular useful for creating an array of images 
   \includegraphics[width=8.4cm]{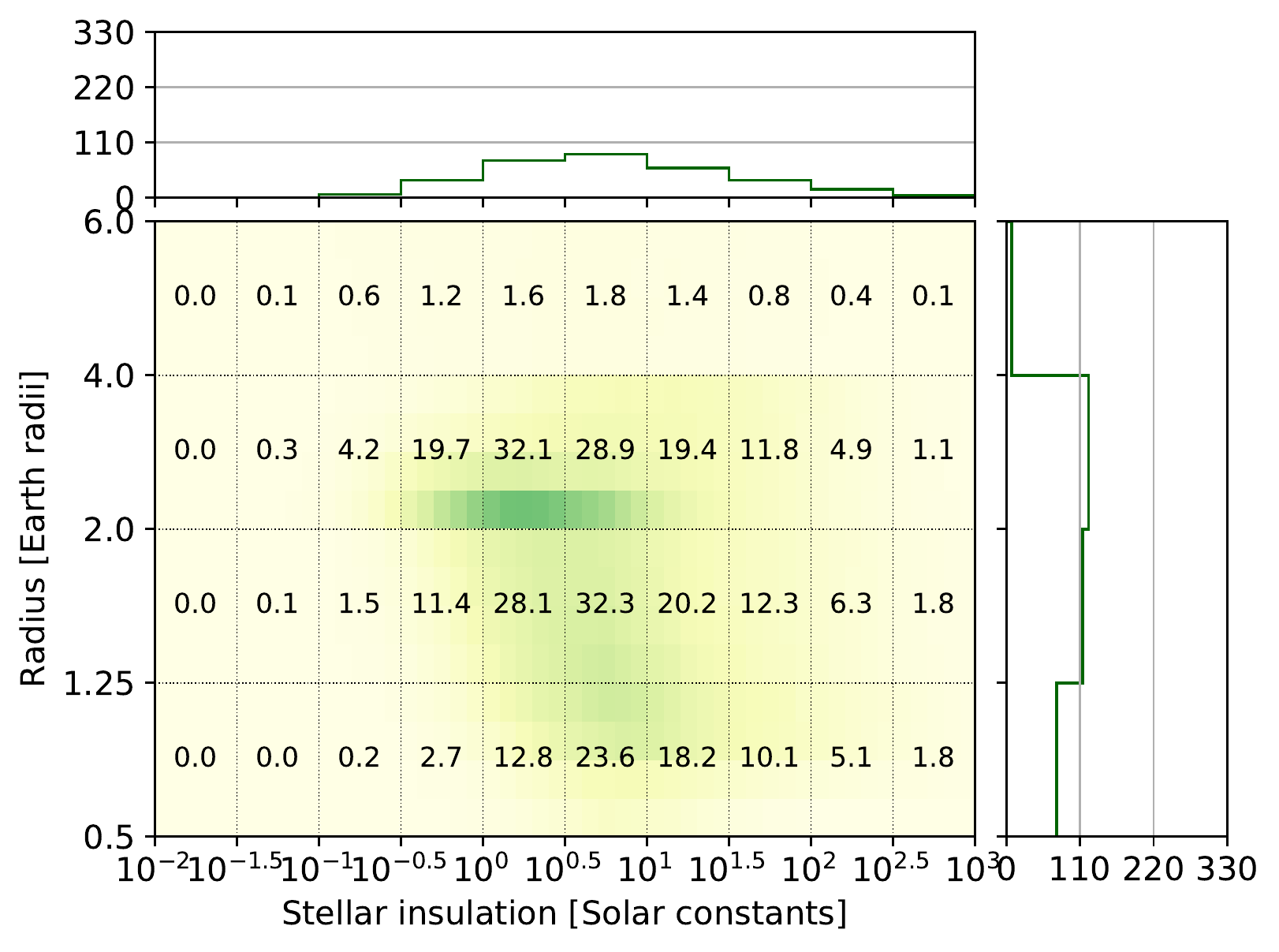}
   \includegraphics[width=8.4cm]{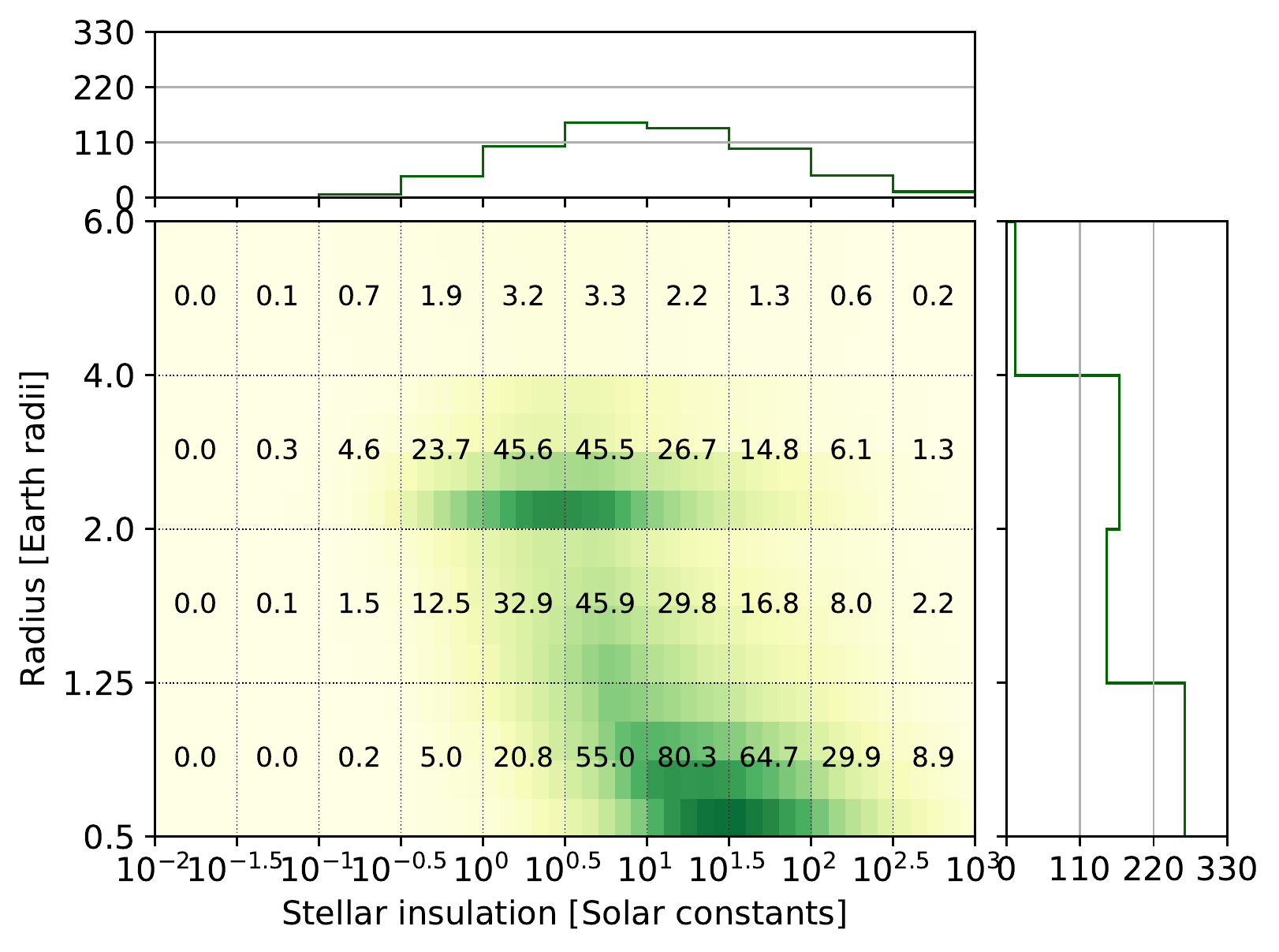}
   \end{tabular}
   \end{center}
   \caption[fig1]  
%>>>> use \label inside caption to get Fig. number with \ref{}
   { \label{fig:1} Number of detectable planets for our space-based MIR nulling interferometer in planet radius vs. stellar insulation space. The expected number of detected planets is given in each grid cell. The left panel assumes the same planet population as Kammerer \& Quanz (2018)\cite{kammerer2018} while the right panel uses the recent results from Hsu et al. (2018)\cite{Hsu2018} for planets orbiting solar-type stars.}
   \end{figure}

Figure~\ref{fig:1} shows the results of our Monte Carlo simulations for our MIR nulling interferometer for both assumed planet populations. It is obvious that Population 2 predicts a significantly higher number of detectable planets in particular for the smallest planets with radii in the range 0.5 -- 1.25 R$_{Earth}$. As planets in this radius range are very likely rocky in composition and not dominated by a significant gas envelope\cite{rogers2015,chen2017} they represent a particular interesting subset of planets that are basically out of reach for in-depth studies with existing technologies. In terms of numbers, in the baseline scenario our nulling interferometer would detect 320 planets in the shown discovery space assuming Population 1 and almost 600 assuming Population 2. 

\begin{figure}[t!]
   \begin{center}
   \begin{tabular}{c} %% tabular useful for creating an array of images 
   \includegraphics[width=8.4cm]{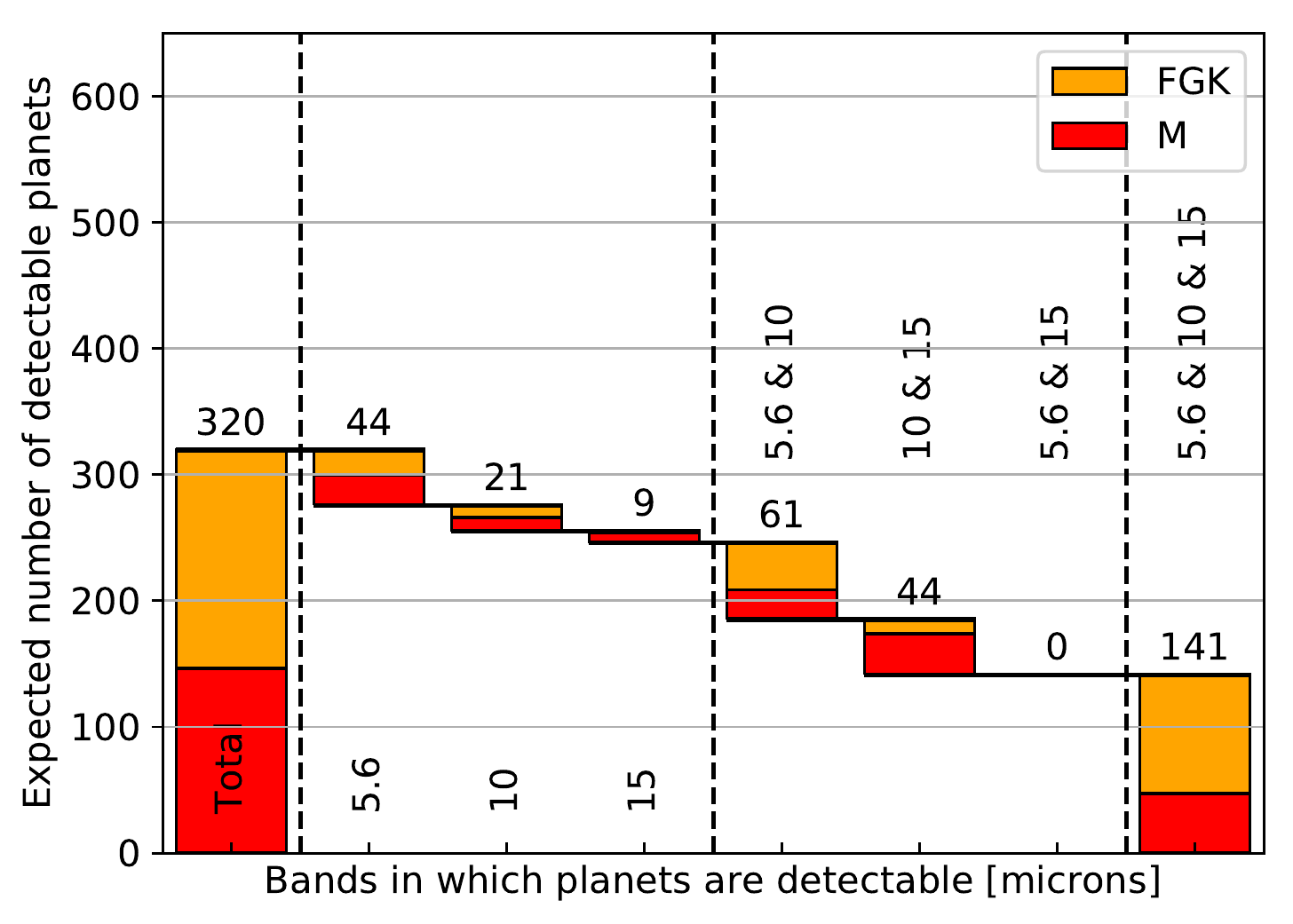}
   \includegraphics[width=8.4cm]{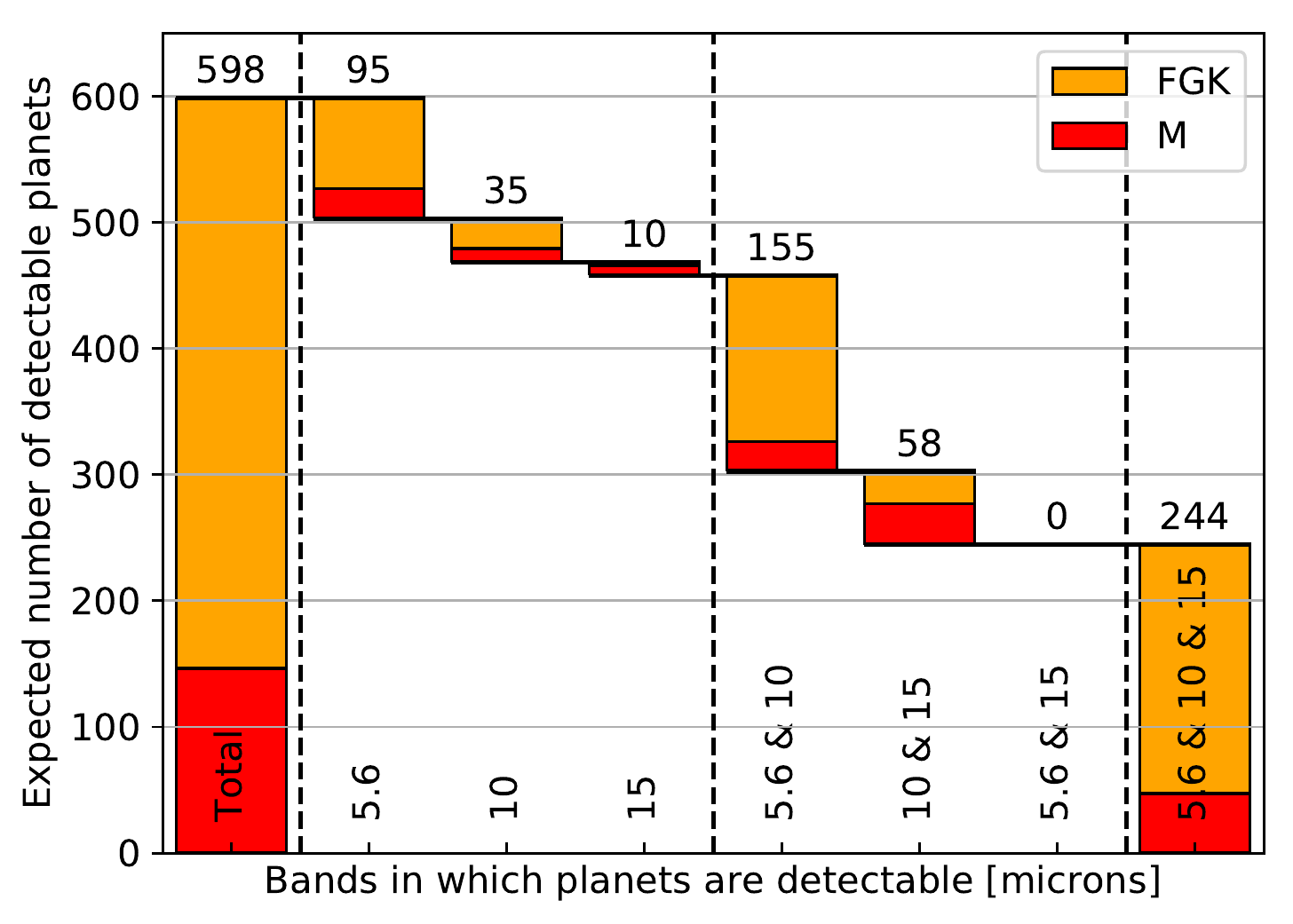}\\
   \includegraphics[width=8.4cm]{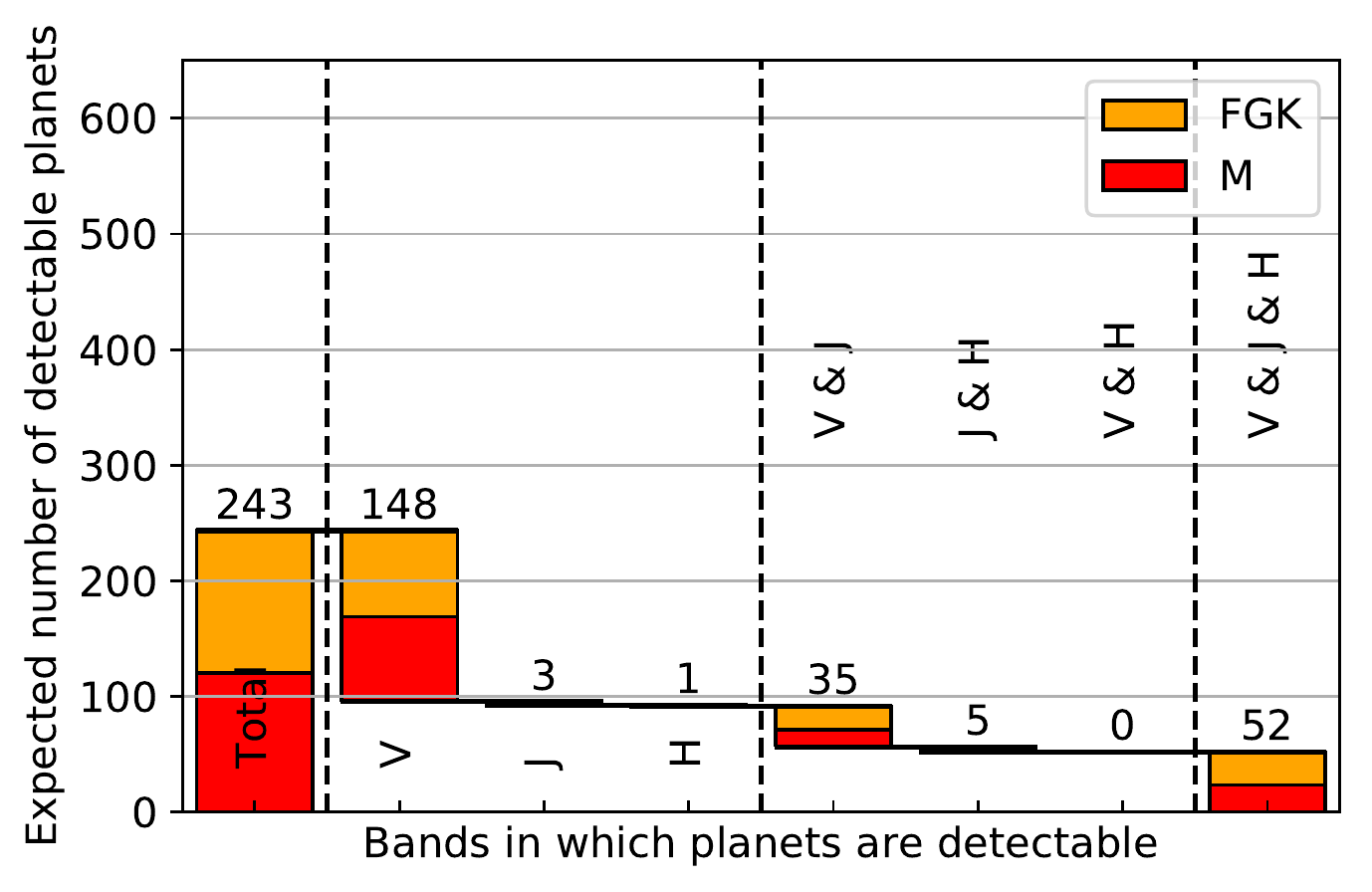}
   \includegraphics[width=8.4cm]{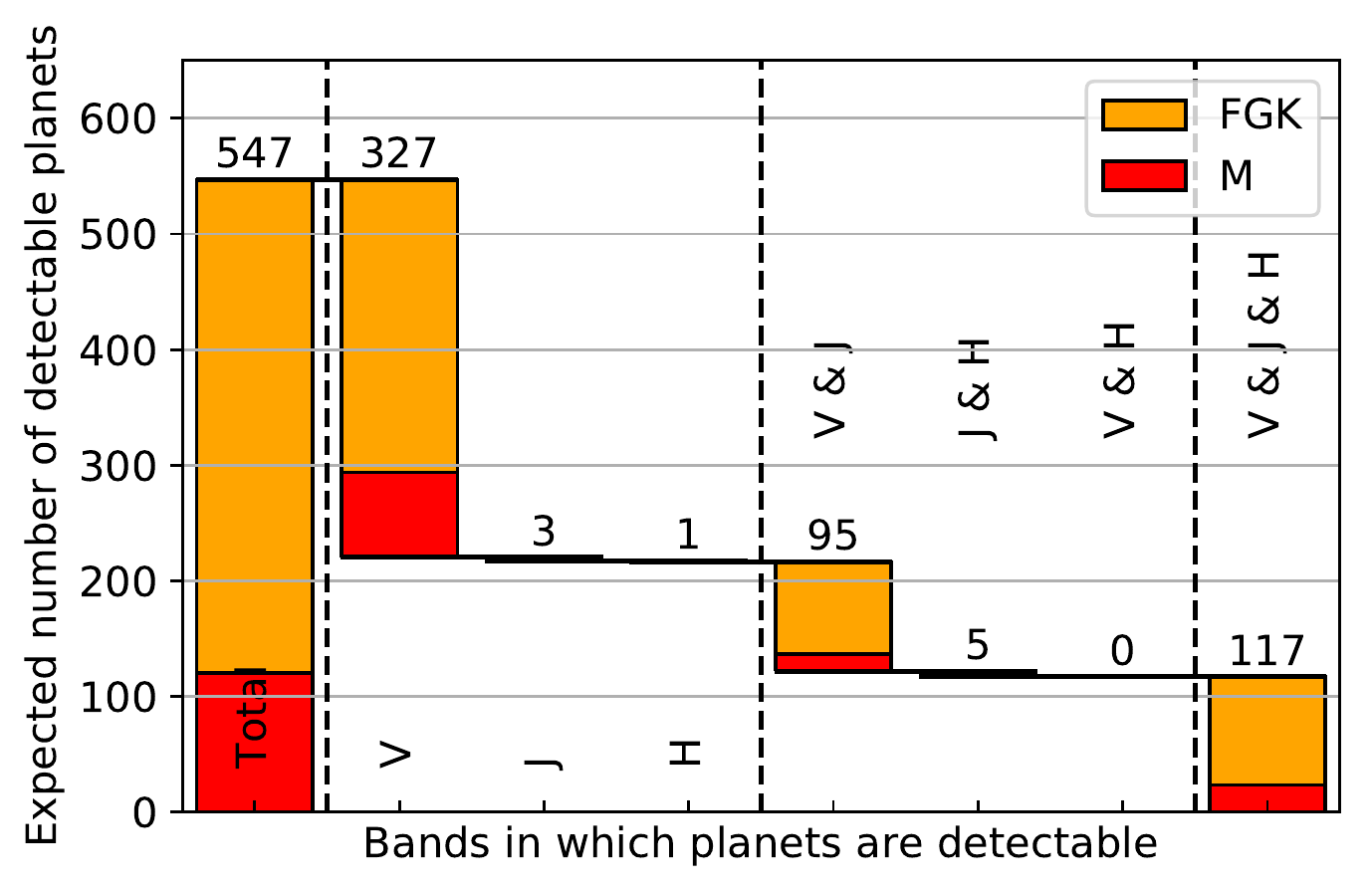}
   \end{tabular}
   \end{center}
   \caption[fig2] 
%>>>> use \label inside caption to get Fig. number with \ref{}
   { \label{fig:2} Breakdown of the total number of detectable planets by spectral type and waveband in which they are detected. Top-left and top-right are for the  baseline scenario of the MIR nulling interferometer assuming Population 1 and 2, respectively. Bottom-left and bottom-right are for the baseline scenario of the optical/NIR telescope assuming Population 1 and 2, respectively.}
   \end{figure}

Figure~\ref{fig:2} (top panels) shows the same results, but now split into spectral types of the host stars (FGK vs. M) and also showing in which bands the planets are detected. For Population 1, the 320 detectable planets are split almost evenly between FGK stars and M stars and 75\% of all planets are detected in at least two of the three bands (5.6, 10 and 15\,$\mu$m). When planets are not detected in one of the bands it can either be that they are too close to the star, and hence no longer spatially resolved at longer wavelengths, or they are too cold and don't emit enough flux at shorter wavelengths. This interpretation is supported by the fact that there is no planet that is detected at 5.6 and 15\,$\mu$m but not at 10 $\mu$m. The top-right panel illustrates again the strong increase in detectable planets around G stars for simulations based on Population 2. The increase is over-proportional in the shorter wavelength range, i.e., for planets orbiting closer to their stars and receiving significantly higher stellar insulation than Earth. 

For comparison, the bottom panels of Figure~\ref{fig:2} show the results for the optical/NIR telescope. In particular for simulations with Population 1 (left panel) the total number of planets detectable in reflected light (243) is significantly smaller than that for planets detected with the MIR nulling interferometer. The split by spectral types is, however, almost identical with half of the planets being detected around M-stars and the other half around FGK stars. $\sim$60\% of the planets are only detected in the V band (0.5 $\mu$m) and only $\sim$20\% in V, J and H (0.5, 1.2, and 1.6 $\mu$m). The reason is simply the less and less favorable inner working angle going from the V band to longer wavelengths. Looking at the results based on Population 2 (right panel) it is striking that here the increase in the number of detectable planets is much stronger for the optical/NIR mission compared to the MIR interferometer. Hence, an optical/NIR mission benefits more, in a relative sense, from the increase in occurrence rate of small planets orbiting Sun like stars in Population 2 compared to Population 1. Still, even for Population 2 the fraction of planets only detectable in the V band  ($\sim$60\%) or detectable in all three bands  ($\sim$20\%) remains unchanged compared to the simulations using Population 1. Whether or not a planet is detected in multiple bands is important for the characterization potential of the mission. 

\subsection{Potential for higher exoplanet yields} 
\begin{figure} [t!]
   \begin{center}
   \begin{tabular}{c} %% tabular useful for creating an array of images 
   \includegraphics[width=5.2cm]{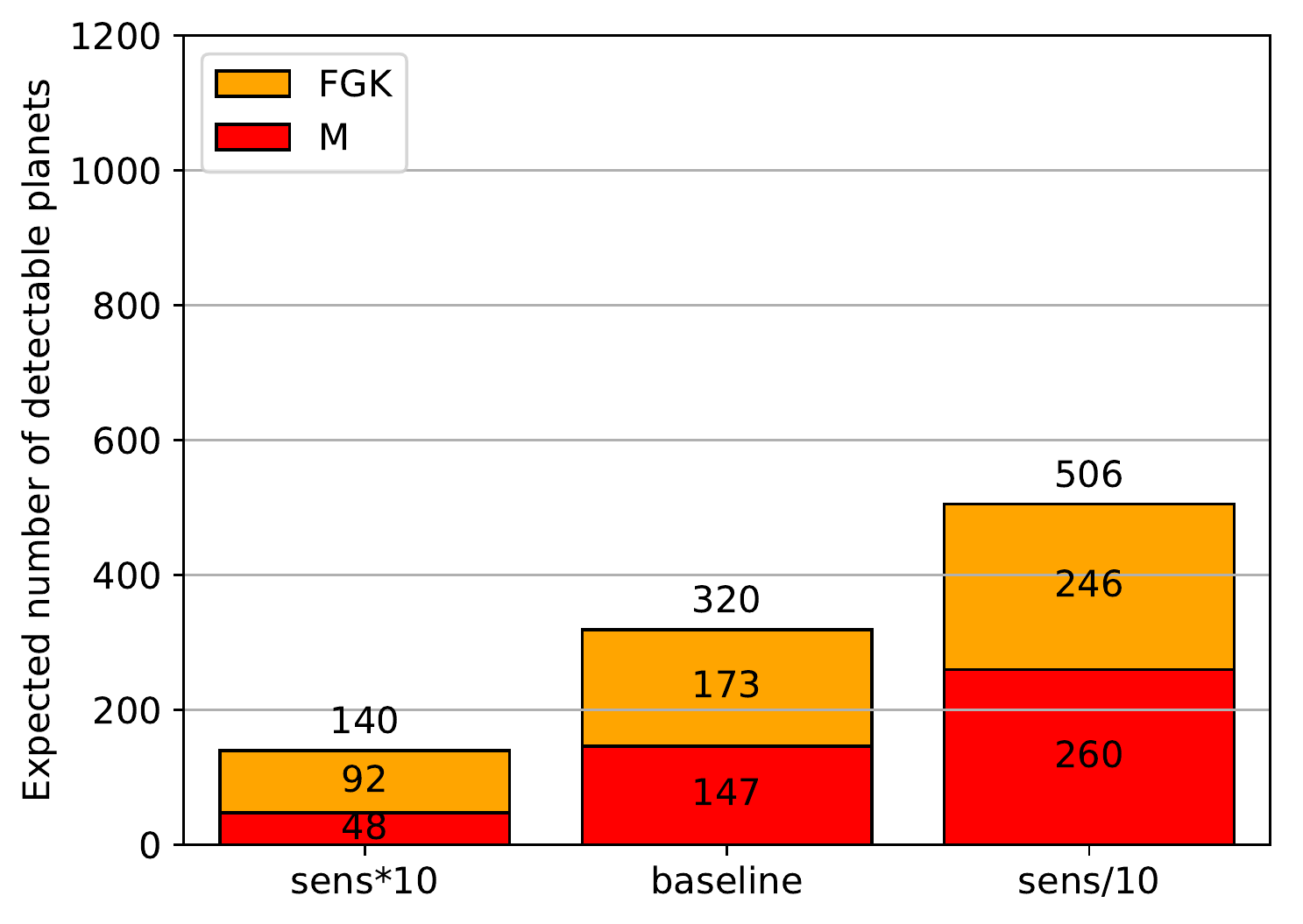}
   \includegraphics[width=5.2cm]{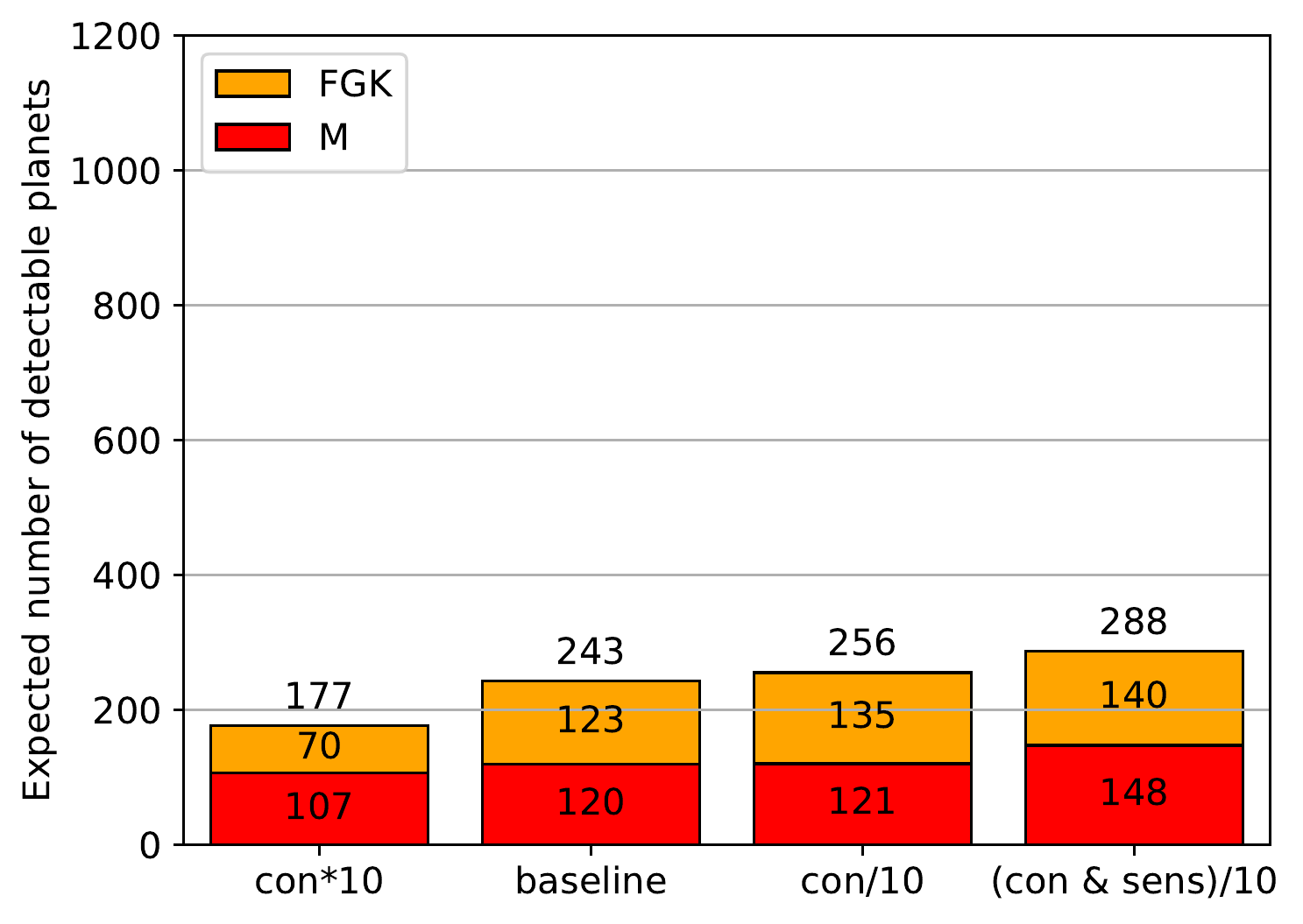}
   \includegraphics[width=5.8cm]{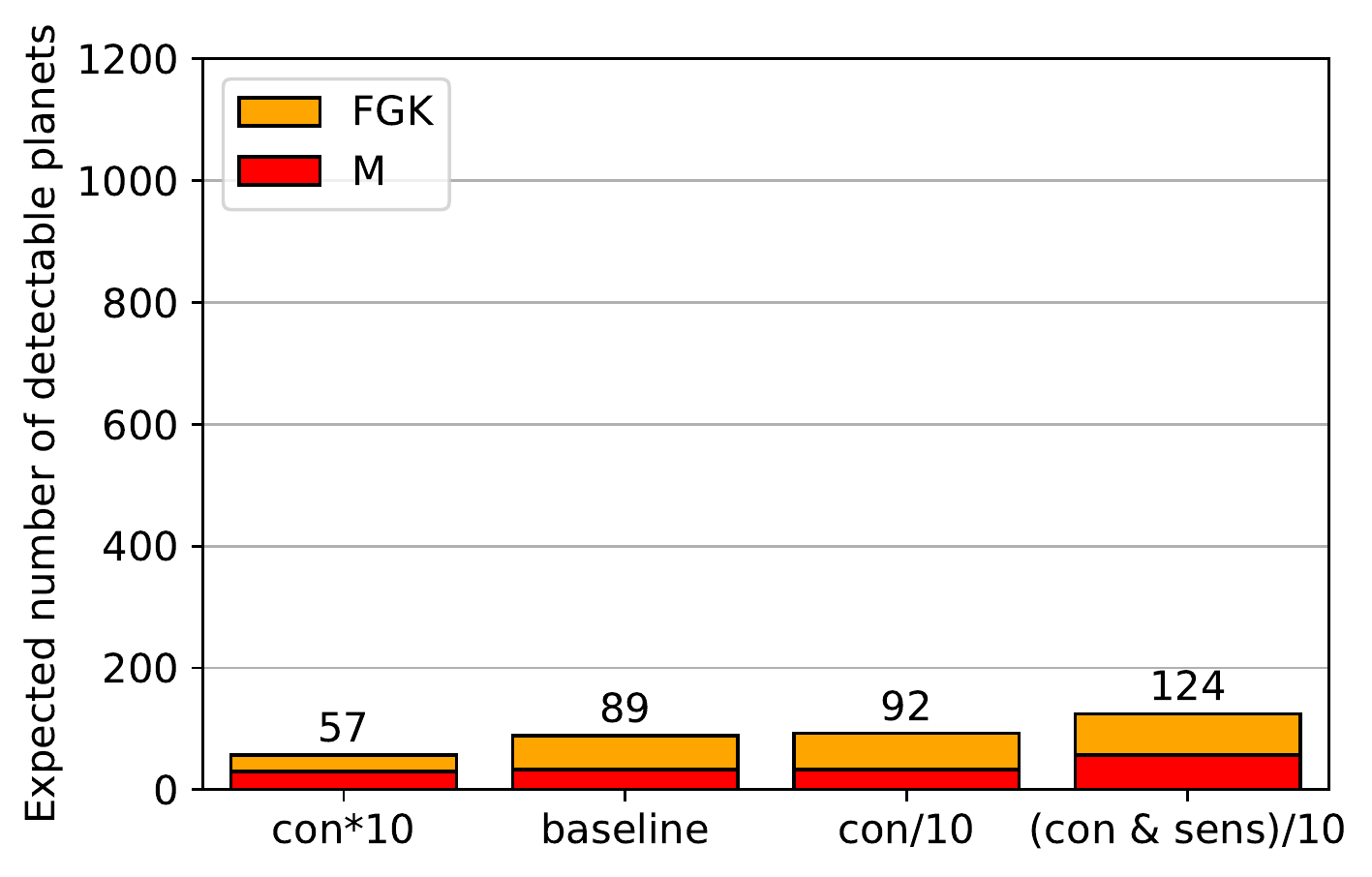}\\
   \includegraphics[width=5.cm]{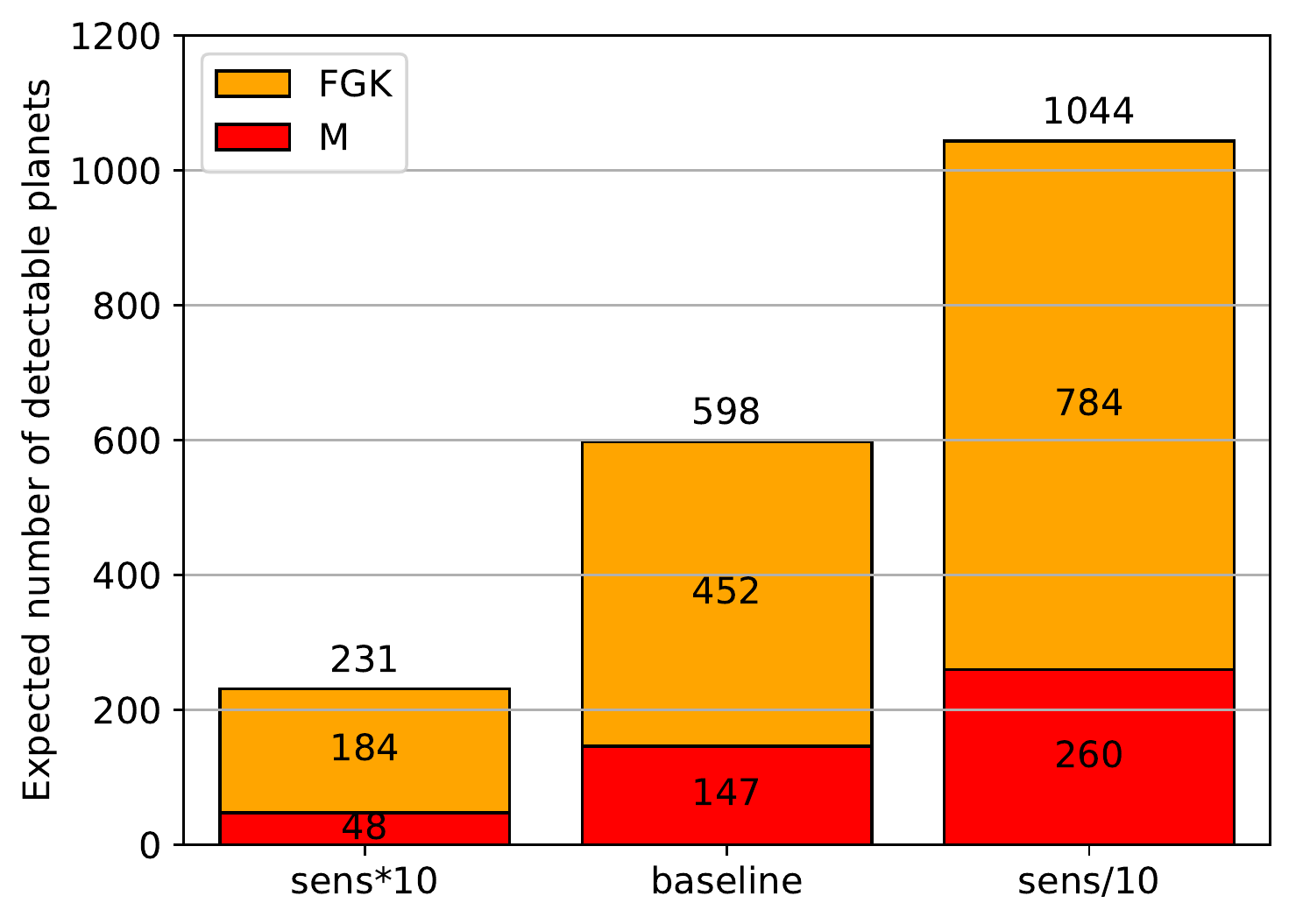}
   \includegraphics[width=5.6cm]{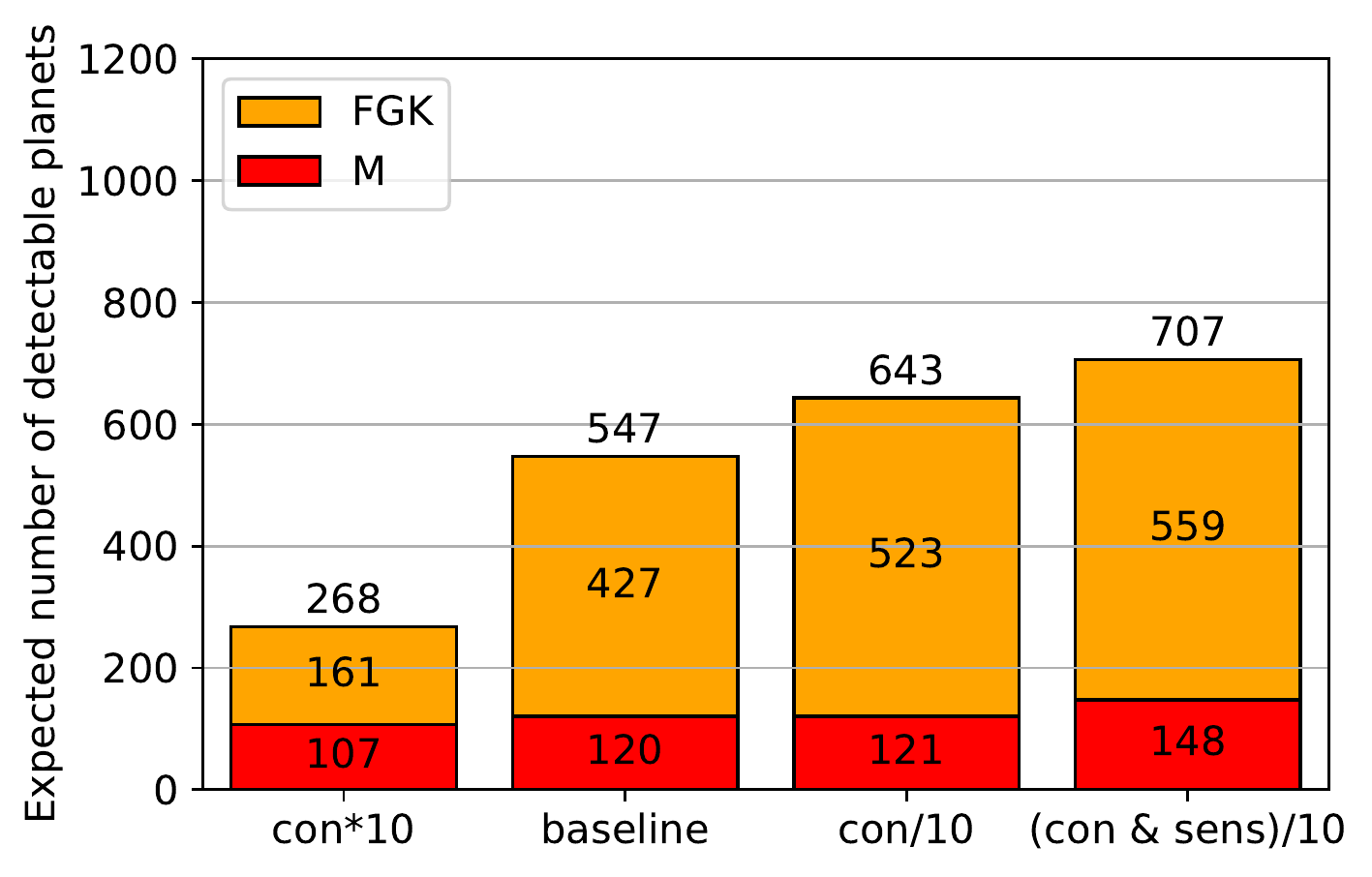}
   \includegraphics[width=5.6cm]{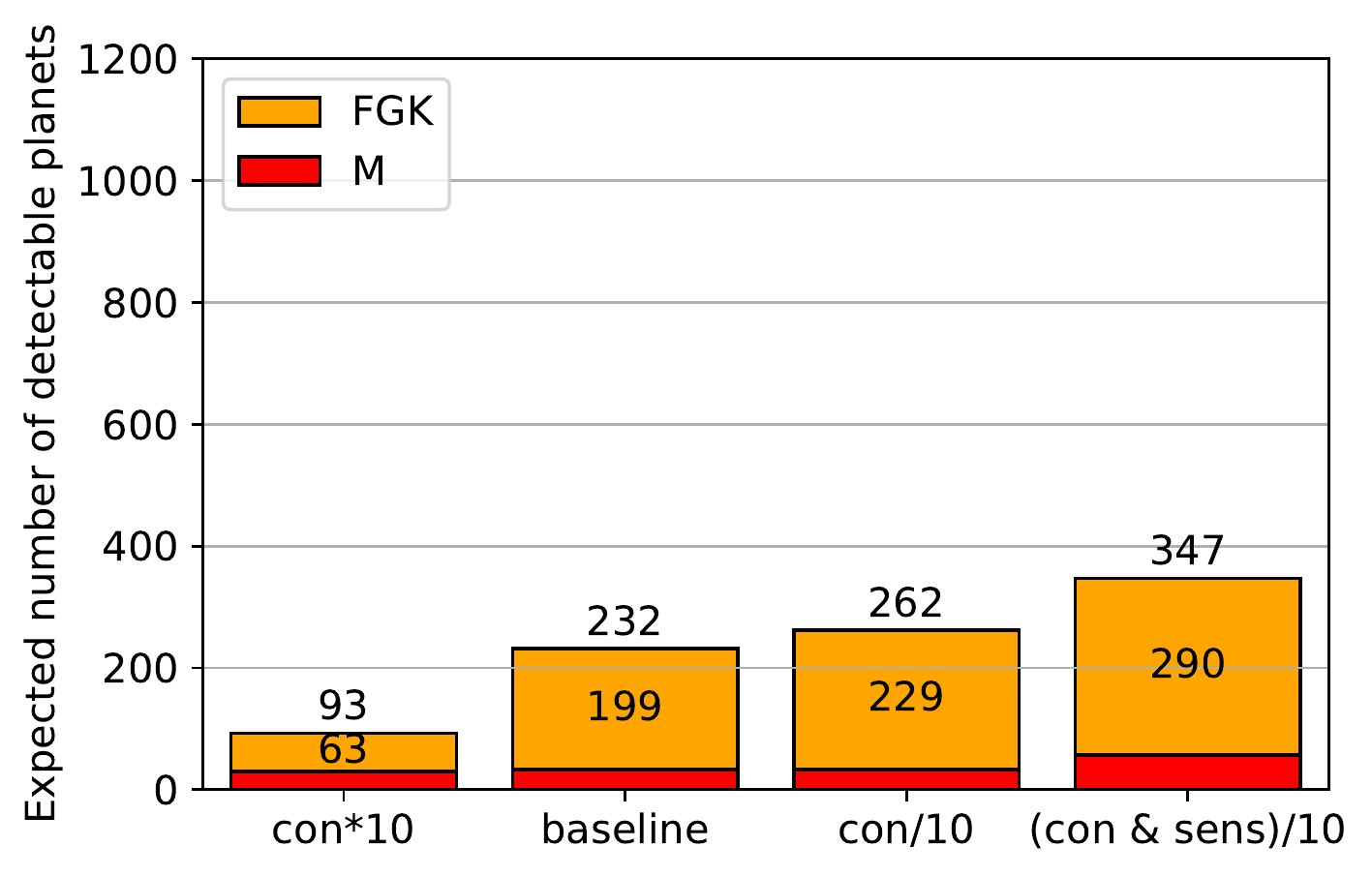}
   \end{tabular}
   \end{center}
   \caption[Fig.3] 
%>>>> use \label inside caption to get Fig. number with \ref{}
   { \label{fig:3} Impact of changing some of the key assumptions on the estimated number of detectable exoplanets. The top row shows the results assuming Population 1, the bottom row the same results assuming Population 2. Left column: results for the MIR nulling interferometer and the impact of assuming a sensitivity that is a factor of 10 better or worse compared to the baseline scenario. Middle column: results for the optical/NIR telescope and the impact of assuming a contrast performance that is a factor of 10 better or worse compared to the baseline scenario, as well as assuming contrast \emph{and} sensitivity performance both to be a factor of 10 better. Right column: same as for the middle column, but now assuming an aperture size of 6 m only instead of 12 m.}
   \end{figure}

It is obvious from Figures~\ref{fig:1}  and~\ref{fig:2} that the assumed underlying population of exoplanets has a significant impact on the planet yield for either mission concept. However, it is also important to quantify the impact of the key technical assumptions. In Figure~\ref{fig:3} we compare the detection yield of the baseline scenarios to that assuming different sensitivities, spatial resolutions and contrast performances. We do that for both mission types and both underlying planet populations. It shows that, not only in an absolute sense, the changes when assuming Population 2 are stronger (as the overall number of planets is higher), but also in a relative sense the changes are more significant. 

Overall the MIR interferometer is quite sensitive to changes in the assumed sensitivity; a factor of ten worse sensitivity cuts the number of detectable planets by more than a factor of two; a factor of ten better sensitivity yields a factor of 1.6 more planet detections. For the optical/NIR telescope the changes in planet yield are rather modest. Increasing the contrast performance by a factor of ten (i.e., to 1e-11) increases the planet detection rate by less than 20\%. However, a contrast a factor of ten worse than the baseline (i.e., 1e-9) reduces the number of planet detections significantly. This shows that a contrast of 1e-10 would  indeed be an important requirement for an optical/NIR mission (cf. Kammerer \& Quanz 2018\cite{kammerer2018}). Increasing the contrast \emph{and} the sensitivity limit at the same time also has only a modest effect (increase $<$20\% for Population 1 and $\sim$30\% for Population 2) compared to the baseline scenario. Hence, the improvement potential for the baseline scenario is rather limited for the optical/NIR telescope. To investigate the impact of the spatial resolution for this mission we also carried out simulations assuming a diameter for the primary mirror of only 6-m (right column in Figure~\ref{fig:3}) in which case we adapted the assumed sensitivities accordingly. As one can see, cutting down the IWA by a factor of 2 compared to the baseline scenario does not cut the total number of detectable planets in half, but leads to comparable losses as the reduction of the contrast performance (at least for Population 1). The relative changes when modifying contrast and/or sensitivity are almost identical for the 12 m and the 6 m primary mirror.

\subsection{Detectability of known exoplanets} 
In addition to quantifying the exoplanet detection yield in a statistical sense, based on the results from the Kepler mission, it is also interesting to estimate which of the currently known exoplanets one could detect with a space-based nulling interferometer. To do so, we accessed the Exoplanet Archieve\footnote{https://exoplanetarchive.ipac.caltech.edu/cgi-bin/TblView/nph-tblView?app=ExoTbls\&config=planets} and computed the expected equilibrium temperatures for the planets listed there using the same approach as in Kammerer \& Quanz (2018)\cite{kammerer2018}. In case the planets were detected via radial velocity measurements, we converted the estimated minimum mass into a radius estimate. For planets with masses $<$7\,M$_{Earth}$ we assumed an Earth-like density, for planets with higher masses a Jupiter-like density, which, for a given mass, determines the radius. A planet is considered detectable if its semi-major axis is larger than the IWA of the interferometer and if its flux -- again approximated via blackbody emission -- is higher than the sensitivity limits. In total, between 40 and 60 planets with radii $<$6\,M$_{Earth}$ could be detect in at least one of the three bands considered here; a significant fraction of that even in all three bands. As these planets and orbital parameters are already known, observations could be scheduled to maximize the chances of detection with the interferometer, which in turn would allow for the determination of the true mass of objects detected via RV measurements. 

This back-of-the-envelope estimate is certainly not perfect and could be improved upon. However, similar as for the Monte Carlo simulations above, it gives a good indication for the number of potential detections. 

\subsection{Quantifying the number of potentially habitable planets} 
One of the key results from the previous sections is that a space-based MIR nulling interferometer, with technical specifications similar to the ones once foreseen for ESA's Darwin mission concept, could yield hundreds of planet detections within a 2-3 year ``discovery'' phase. Already today the diversity in planetary properties and in their systems architecture is striking and in case one wants to investigate whether this diversity also continues for atmospheric properties\footnote{In our Solar system we have three types of atmospheres: H/He dominated (gas \& ice giants), CO$_2$ dominated (Venus \& Mars), and N$_2$ dominated (Earth \& Titan).} and how this diversity could be linked to planetary formation and evolutionary models one needs indeed a sizable sample of (small) planets orbiting different types of host stars.

In addition to the statistical aspect of exoplanetology that such a mission would have, the second driving aspect concerns the potential habitability of detectable planets. How many potentially habitable (or even inhabited?) planets could be identified during the ``discovery'' phase so that during the follow-up ``characterization'' phase chances of finding indeed habitable planets are maximized? To do that we focused our analysis on a subset of detectable planets: those with radii 0.5 R$_{Earth}\,<\, $R$\,<\,$1.75 R$_{Earth}$ and experiencing stellar insulations S between 0.5\,$<$\,S/S$_0$\,$<$\,1.5 with S$_0$ denoting the solar constant. This range in energy flux describes roughly the range of the empirical habitable zone in the Solar system\cite{kaltenegger2017}, where the inner edge is referred to as ``Recent Venus'' limit and the outer edge as ``Early Mars'' limit. As for the radius range our choice is motivated by the idea that also planets with a non-negligible H/He envelope, i.e., planets with slightly larger radii than $\sim$1.5 R$_{Earth}$\cite{rogers2015,chen2017}, might be habitable and even biomarkers could be detected\cite{seager2013}.

\begin{figure} [t!]
   \begin{center}
   \begin{tabular}{c} %% tabular useful for creating an array of images 
   \includegraphics[width=5.6cm]{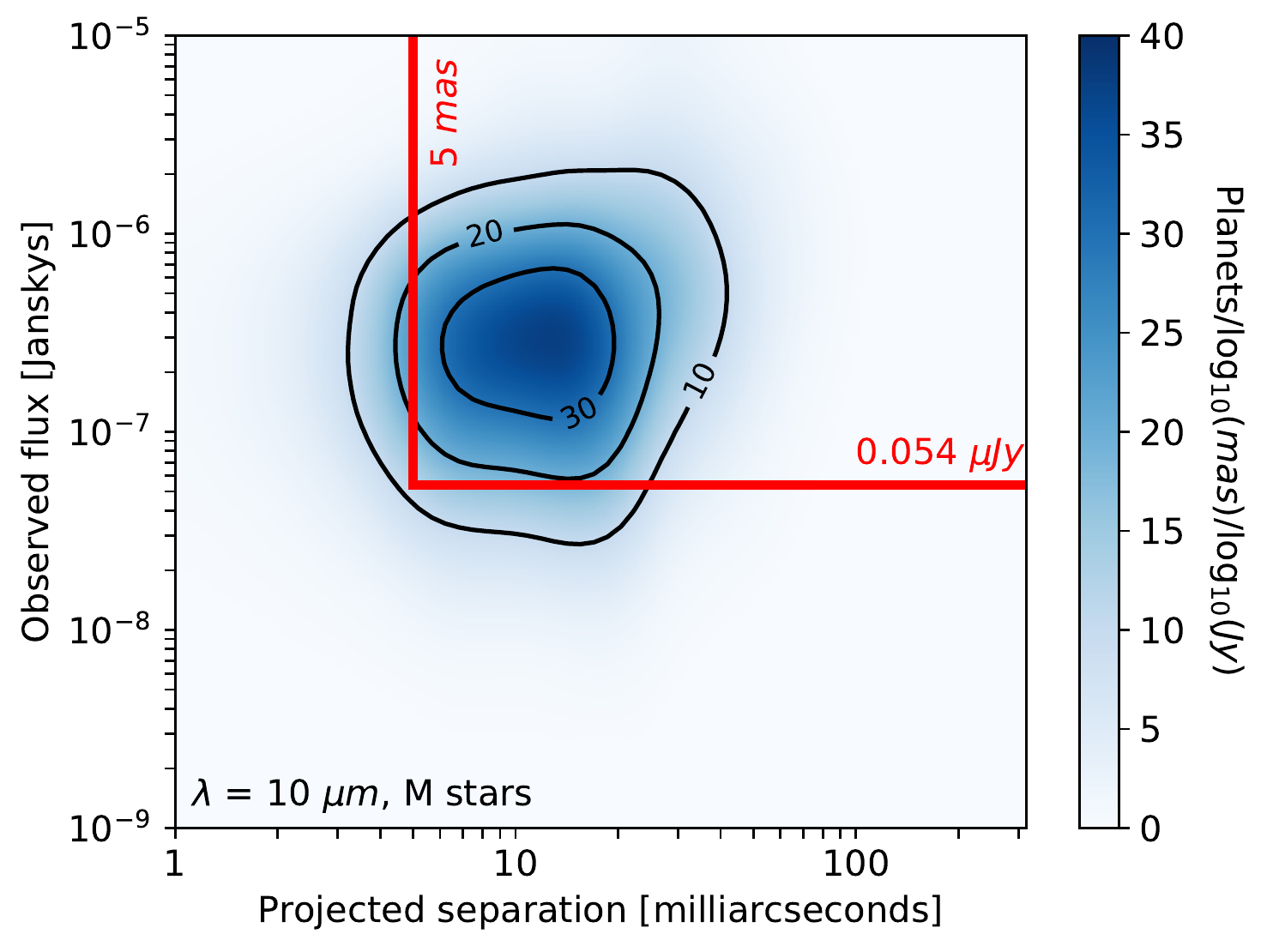}
   \includegraphics[width=5.6cm]{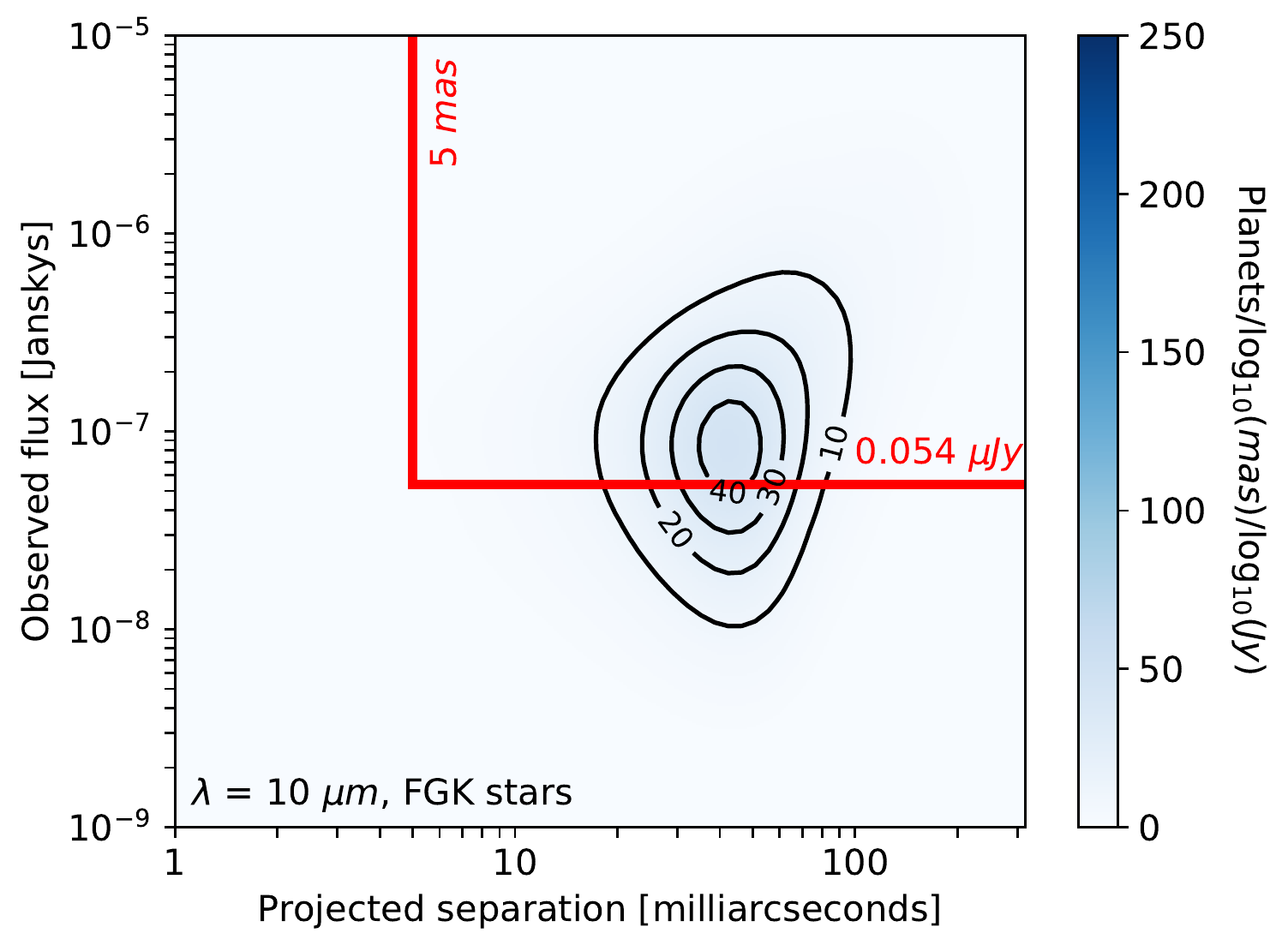}
   \includegraphics[width=5.6cm]{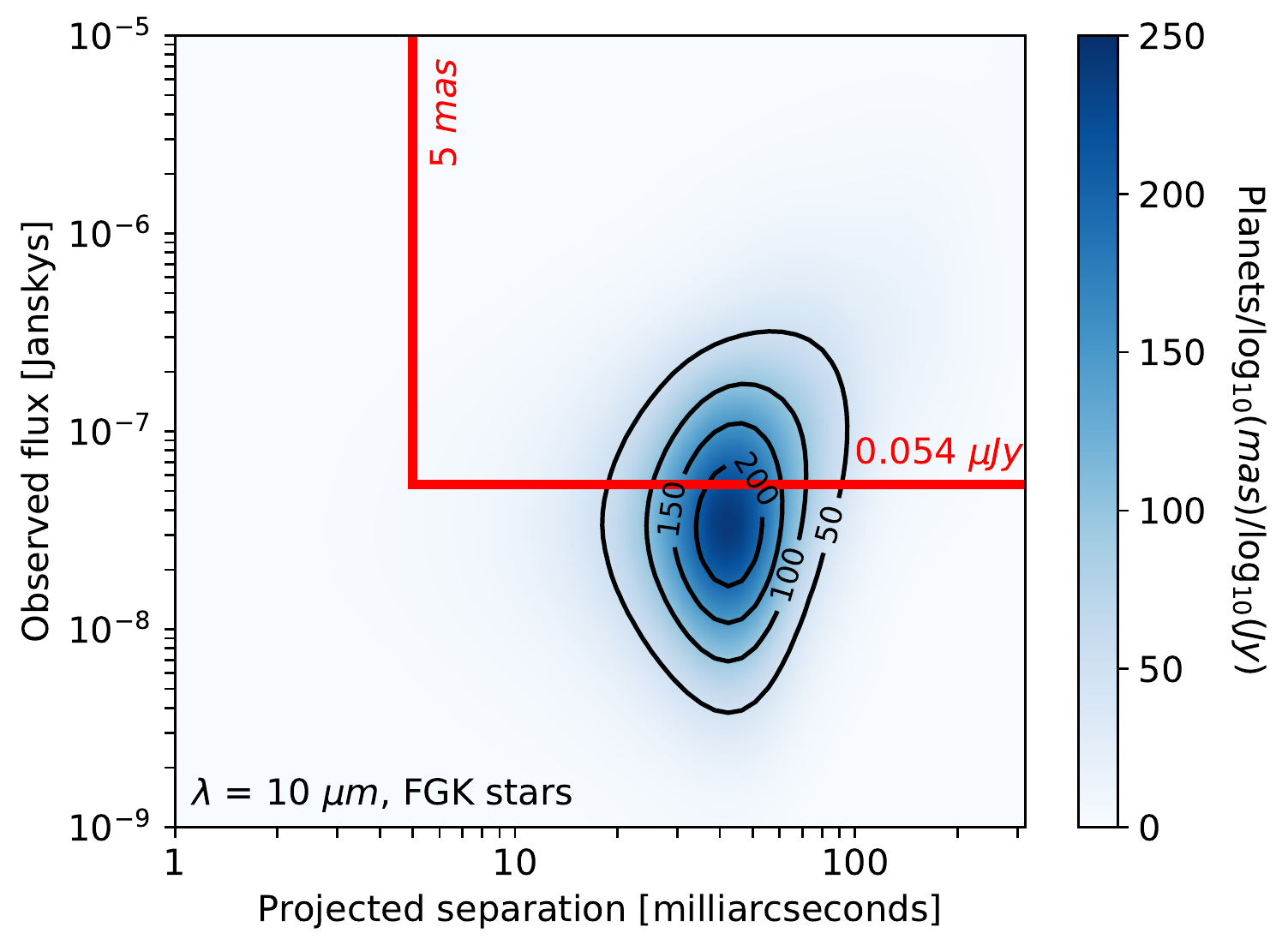}
   \end{tabular}
   \end{center}
   \caption[Fig.4] 
%>>>> use \label inside caption to get Fig. number with \ref{}
   { \label{fig:4} Contour plots showing where the bulk of detectable planets with radii R between 0.5 R$_{Earth}\,<\, $R$\,<\,$1.75 R$_{Earth}$ and insulations S between 0.5\,$<$\,S/S$_0$\,$<$\,1.5 sit in the flux-separation plane. The red lines indicate the assumed IWA at 10\,$\mu$m wavelength and a sensitivity limit that is a factor of 10 better than in the baseline scenario. From left to right the different panels are for M stars, FGK stars (Population 1), and FGK stars (Population2), respectively.}
   \end{figure}

It turns out that for this particular subset of detectable planets, higher sensitivity is important. In Figure~\ref{fig:4} we plot the population of detectable exoplanets in the flux-separation plane (for M stars and both populations of FGK stars) and overplot the IWA and a ten times better sensitivity limit as assumed in the baseline scenario. It shows that while the IWA is sufficient to detect the bulk of exoplanets in the subset defined above even for M stars -- where the planets have to be very close to their host stars to receive sufficient insulation -- a higher sensitivity than assumed in the baseline scenario is needed to probe the bulk, or at least a significant fraction, of planets around FGK stars. This has to do with the fact that, on average, FGK stars are farther away than M stars requiring longer integration times for the planets to be detected. The apparent difference between the middle (Population 1) and right-hand (Population 2) panels in Figure~\ref{fig:4} can be explained by looking at Figure~\ref{fig:1}: Population 2 has a significantly higher number of smaller planets compared to Population 1, which shifts the bulk of the population in Figure~\ref{fig:4} downwards. 

\begin{figure} [t!]
   \begin{center}
   \begin{tabular}{c} %% tabular useful for creating an array of images 
   \includegraphics[width=8.4cm]{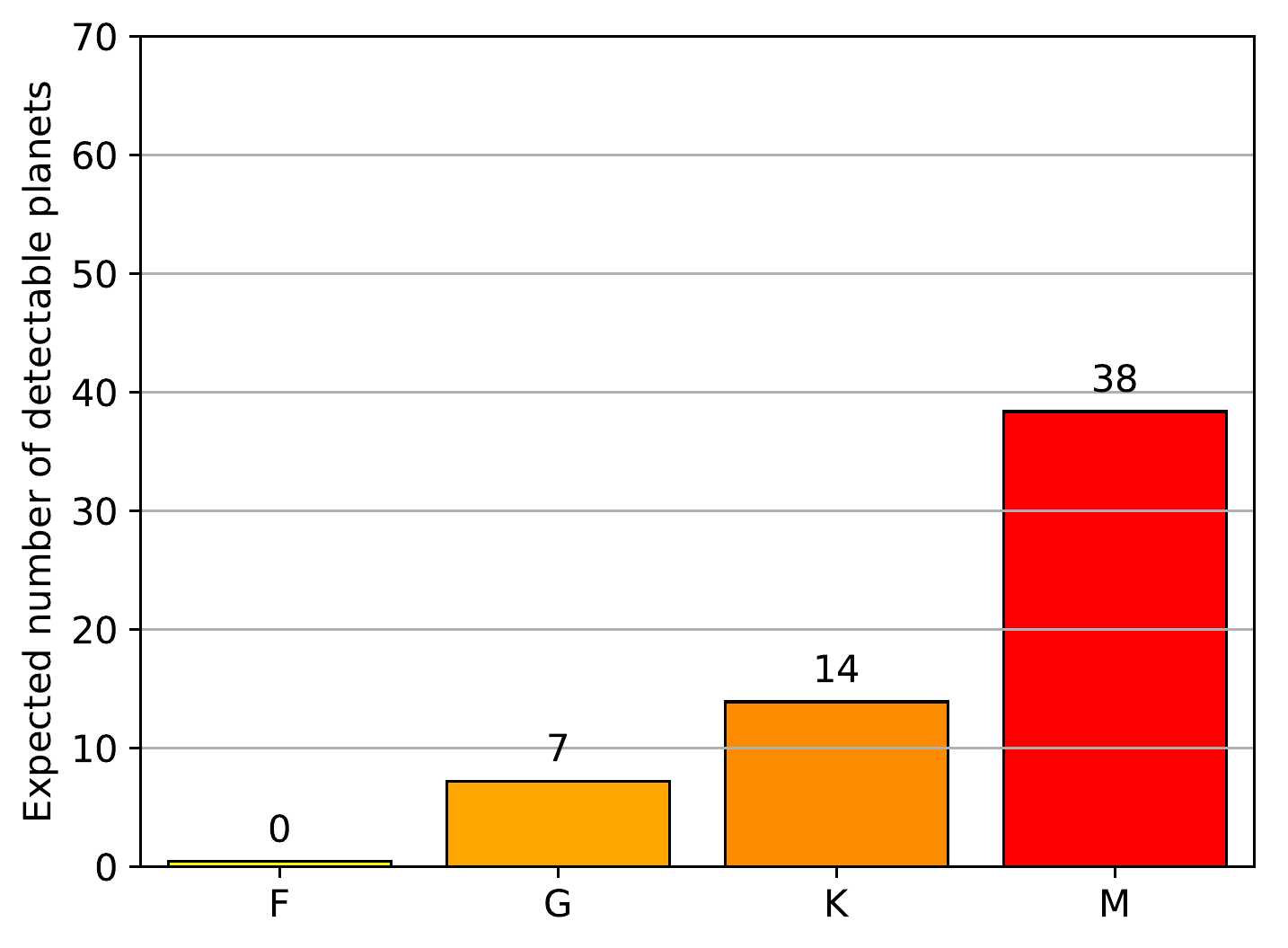}
   \includegraphics[width=8.4cm]{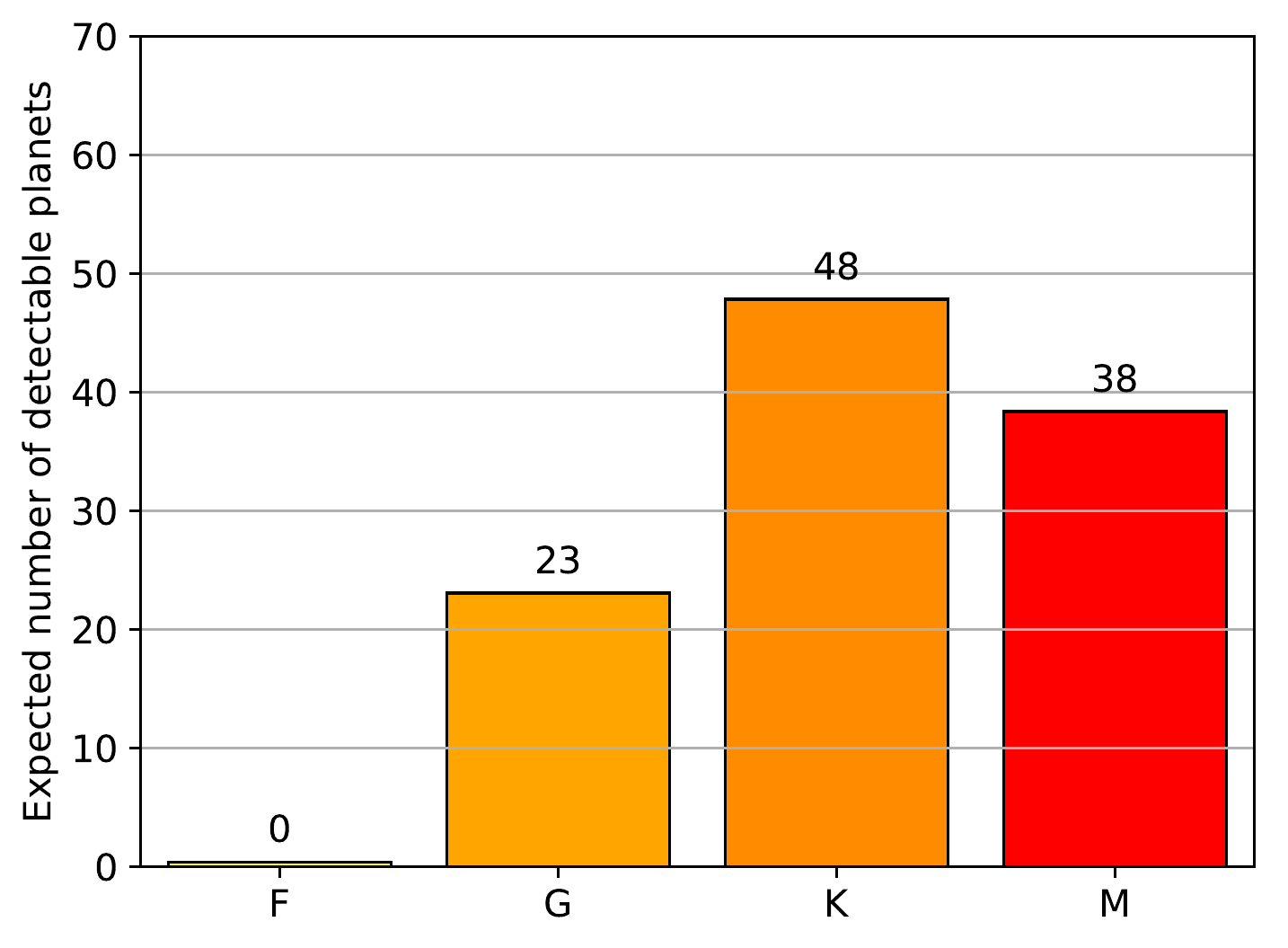}
   \end{tabular}
   \end{center}
   \caption[Fig.5] 
%>>>> use \label inside caption to get Fig. number with \ref{}
   { \label{fig:5} Total number of detectable planets at 10\,$\mu$m wavelength with radii R between 0.5 R$_{Earth}\,<\, $R$\,<\,$1.75 R$_{Earth}$ and insulations S between 0.5$<$S/S$_0$$<$1.5 as a function of spectral type of their host stars and assuming the baseline IWA, but a factor 10 better sensitivity. The left panel is for Population 1 and right panel for Population 2. }
   \end{figure}

Figure~\ref{fig:5} summarizes this analysis in terms of total number of planet detections. Several dozens of exoplanets in this interesting range of parameter space are -- in principle -- accessible with a space-based MIR nulling interferometer. This figure also illustrates again the significant difference between the two planet populations making the results using Population 2 particularly promising. Overall, the relative increase in detectable planets in the subset of parameter space considered here is the same as for the overall planet population shown in Figure~\ref{fig:3}. It is worth noting that for F stars no planet in the defined range of parameter space seems detectable, which we attribute to the too high flux levels planets experience around these stars for the period distribution we consider.   

Another approach to try and quantify how many detectable planets could be habitable is by linking the publicly available {\tt HUNTER} code package\cite{zsom2015} with the results from our Monte Carlo simulations. While in the analysis described above, stellar insulation and planet radius were the only two parameters considered, the {\tt HUNTER} code is based on a large grid of exoplanet atmospheres with several key parameters that can influence planetary habitability (such as atmospheric composition, relative humidity, and surface pressure). Even though none of these parameters are explicitly known, {\tt HUNTER} allows for a habitability assessment in a statistical sense by quantifying how likely it is that a planet (with given insulation and radius) has a surface temperature amendable to liquid water. The details of the {\tt HUNTER} code package as well as the assumed priors for the various parameters are explained in detail in Zsom (2015)\cite{zsom2015}. For our analysis we did not modify the baseline settings of {\tt HUNTER}.

\begin{figure} [b!]
   \begin{center}
   \begin{tabular}{c} %% tabular useful for creating an array of images 
   \includegraphics[width=8.4cm]{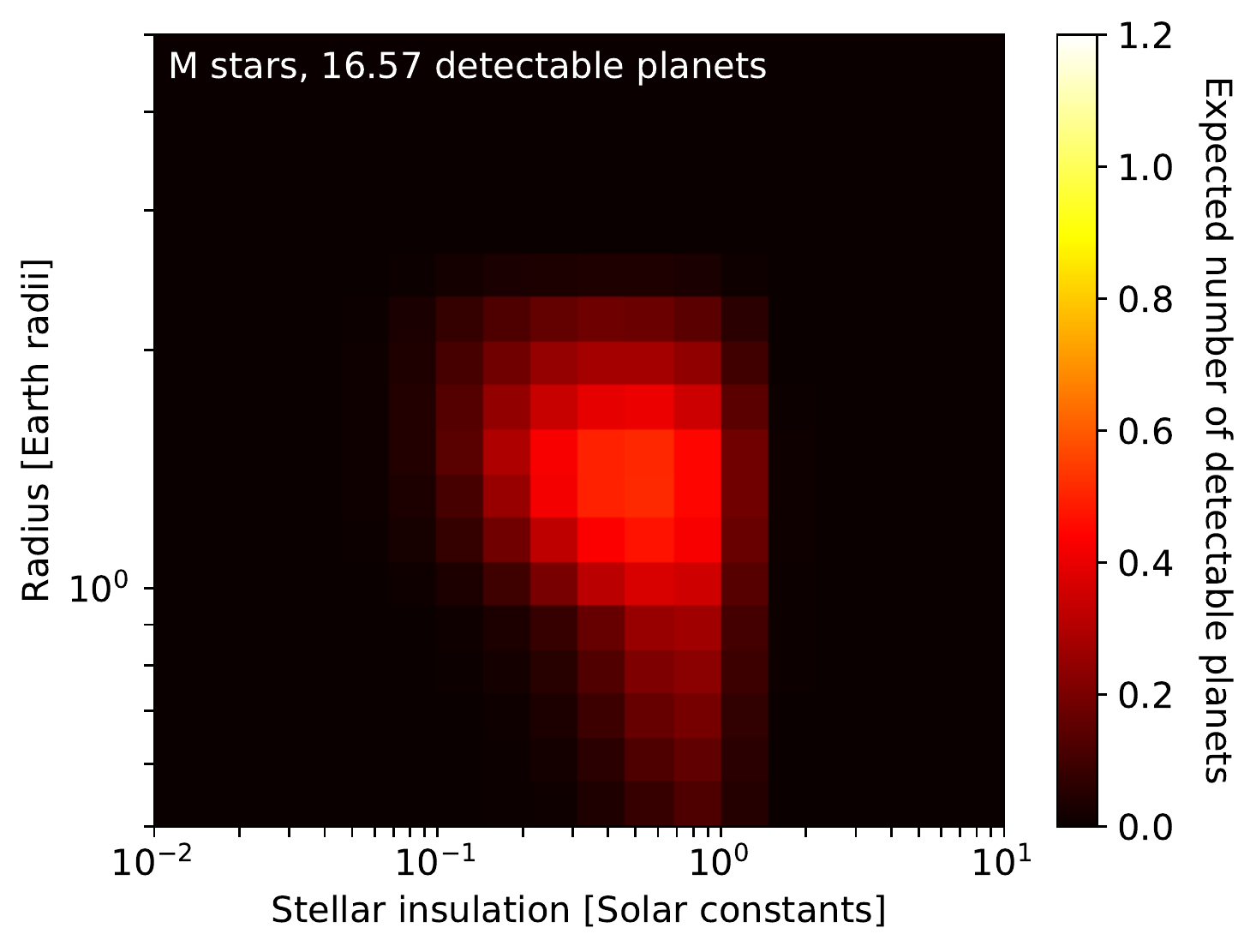}
   \includegraphics[width=8.4cm]{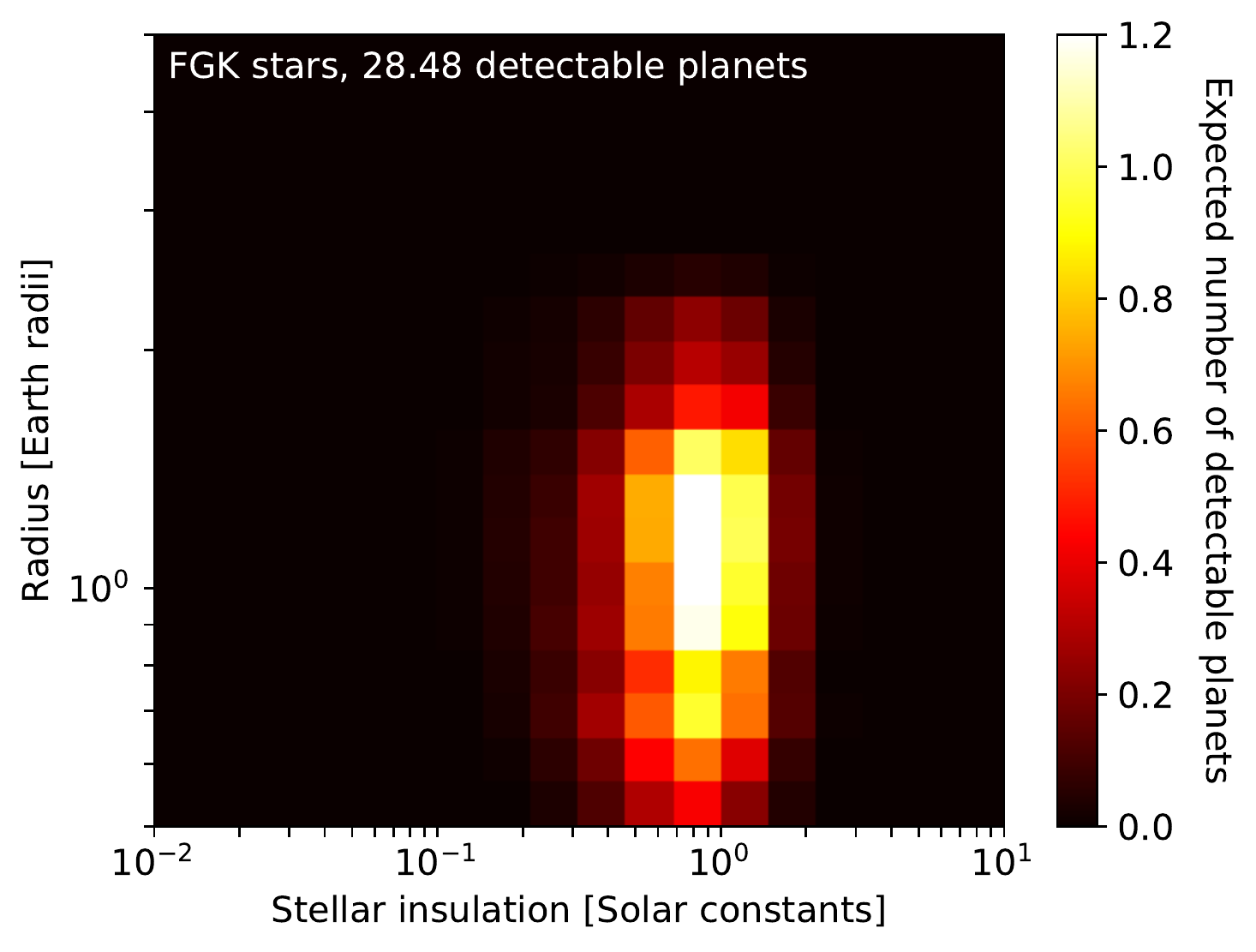}
   \end{tabular}
   \end{center}
   \caption[Fig.6] 
%>>>> use \label inside caption to get Fig. number with \ref{}
   { \label{fig:6} Total number of detectable habitable planets with a MIR nulling interferometer coupling the {\tt HUNTER} package to the population of planets created in our Monte Carlo simulations. For these plots we assumed the baseline IWA, but a factor 10 better sensitivity, and Population 2. The left panel shows the results for M stars ($\sim$16--17 habitable planets expected), while the right panel shows the results for FGK stars ($\sim$28--29 habitable planets expected).
   }
   \end{figure} 

The results are shown in Figure~\ref{fig:6}. Overall the numbers of potentially habitable planets are lower ($\sim$16--17 around M-stars and $\sim$ 28--29 around FGK stars) than in the analysis shown in Figure~\ref{fig:5}, but it needs to be kept in mind that now additional atmospheric parameters to assess habitability are taken into account. If instead of Population 2 (as shown here) Population 1 were assumed the number of detectable habitable planets around FGK stars would decrease by a factor of three in accordance with the results in Figure~\ref{fig:5}. The results shown here also indicate that there could be habitable planets detected outside of the parameter space considered above -- at least in the framework described by {\tt HUNTER}: even if unlikely planets slightly larger than 1.5 R$_{Earth}$ or receiving less insulation than 0.5 S$_0$ could be habitable. This is a direct result from {\tt HUNTER} for which Zsom (2015)\cite{zsom2015} demonstrated the rather broad region of parameter space where habitable planets could exist. 

One of the key aspects that needs further investigation in the context of the results described in this subsection is the sensitivity that is assumed: the baseline sensitivity only allows for the detection of a comparatively small number of potentially habitable planets, in  particular around G stars. Increasing the integration time per target by a factor of ten is hardly feasible as this would exceed the total mission lifetime (initially planned to be 5 years) by a factor of a few. However, as already mentioned above and indicated in Figure~\ref{fig:4}, not all stars necessarily require a factor of ten increase in integration time to increase the number of detectable habitable planets. As already pointed out in Kammerer \& Quanz 2018\cite{kammerer2018} assuming the same on-source time for each star during the ``discovery phase'' is far from ideal and optimizing the individual on-source times, taking into account the stars' spectral types and distances, is needed to find the right trade-off between the duration of the ``discovery phase'' and the number of potentially habitable planets\cite{lay2007,stark2015}.

\section{THE CHARACTERIZATION PHASE}
\label{sec:results_charac}
The previous section dealt with the number of exoplanets that could in principle be detected with MIR nulling interferometer during a 2-3 year ``discovery phase''. These discoveries would then need to be prioritized for follow-up observations to obtain high SNR spectra with a certain spectral resolution over a certain MIR wavelength range. In the past, estimates what spectral resolution and wavelength coverage were required had been based on atmospheric forward models trying to quantify what spectral features could be detected and measured under certain assumptions. Most studies typically considered a wavelength range between 6 -- 25\,$\mu$m and a spectral resolution of R$\sim$20 to be sufficient\cite{cockell2009,desmarais2002}. Today, a complementary -- and from a data analysis and interpretation perspective more robust -- approach would be to invoke spectral retrieval techniques that allow for a statistical assessment of individual atmospheric parameters. The impact of varying the SNR, the wavelength coverage, or the spectral resolution (or combinations thereof) on the final interpretation of the measurements could then directly be quantified. First analyses in this direction have been applied to Earth twin planets observed in thermal emission\cite{vonparis2013} and reflected light\cite{feng2018}. Broadening the scope of these analyses, e.g., varying atmospheric composition, including key biosignatures and clouds, in a systematic way is beyond the scope of this paper, but would be important to help define the science requirements for the ``characterization phase''. 

%Analyze and discuss some of the spectra from Daniel for habitable planets ($T_surface = 288 K$) for different stars and atmospheres. 

\subsection{Wavelength range requirements}
The 6 -- 25\,$\mu$m wavelength range features absorption bands of key molecules for the characterization of (terrestrial) planetary atmospheres such as O$_3$, N$_2$O, CH$_4$, CO$_2$, and H$_2$O  \cite{desmarais2002,kaltenegger2017}. In particular a combined detection of O$_3$ and CH$_4$ would be  considered a strong indication of biological activity, and also N$_2$O is an promising  biosignature. It's worth mentioning that Earth present-level CH$_4$ is probably difficult to detect at NIR wavelengths and N$_2$O has no detectable features at optical/NIR wavelengths. In principle, the MIR also allows for the estimation of the radius of exoplanets from their bolometric flux and the surface temperature can be probed in case cloud-free, atmospheric windows exist. 

It is challenging to scientifically justify the optimal long wavelength cut-off of a MIR mission. The $\sim$15\,$\mu$m CO$_2$ band should certainly be included and being able to probe the broad water band centered around $\sim$20.5\,$\mu$m could also be useful. Hence, a cut-off at around $\sim$25\,$\mu$m is a good choice.\footnote{The wavelength range of JWST/MIRI extends to $\sim$28$\mu$m.} At the short-wavelength end, it may make sense to go beyond the 6\,$\mu$m that were initially foreseen. It was recently suggested that collision-induced absorption by (N$_2$)$_2$ was detected in disk-integrated spectra of Earth at $\sim$4.15\,$\mu$m in the wing of the $\sim$4.3\,$\mu$m CO$_2$ band\cite{schwieterman2015}. For some terrestrial exoplanets quantifying N$_2$ could provide a means of determining bulk atmospheric composition. Furthermore, looking at the results from Figure~\ref{fig:1}, it seems clear that most exoplanets to be detected will presumably be much warmer than Earth, shifting their peak emission to shorter wavelengths. Already for Earth the transition between where the emission is dominated by scattered light or thermal emission occurs roughly at 3\,$\mu$m\cite{robinson2018}. Given that we have to be up for surprises when starting the characterize the atmospheres of (warm) terrestrial planets, extending the wavelength range down to 3\,$\mu$m needs to be considered. 

\begin{figure} [t!]
   \begin{center}
   \begin{tabular}{c} %% tabular useful for creating an array of images 
   \includegraphics[width=13.5cm]{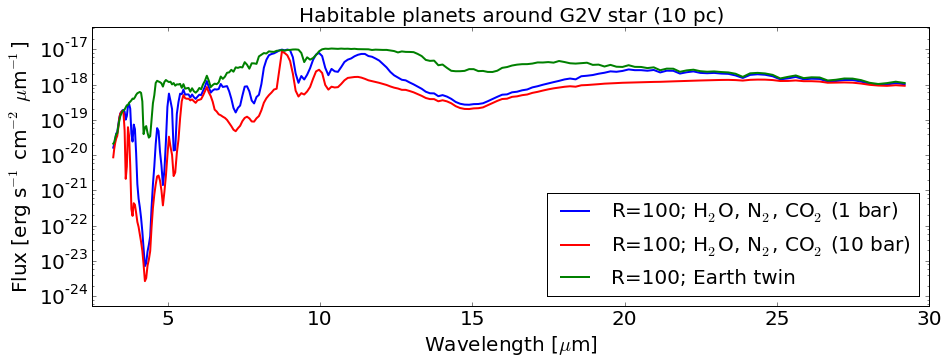}\\
   \includegraphics[width=13.5cm,angle=0]{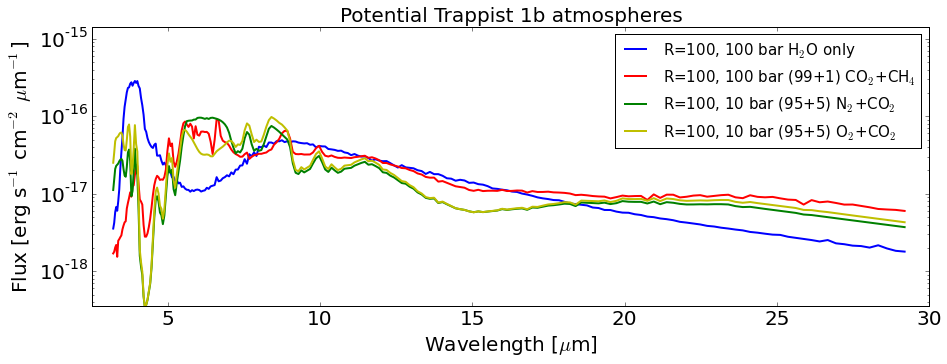}
   \end{tabular}
   \end{center}
   \caption[Fig.7] 
%>>>> use \label inside caption to get Fig. number with \ref{}
   { \label{fig:7} MIR thermal emission spectra of terrestrial planet atmospheres with spectral resolution of R$=$100. Top: Earth-like habitable planets around a solar type star at 10 pc (green: Earth twin atmosphere; blue: N$_2$--CO$_2$--H$_2$O only atmosphere with CO$_2$ surface partial pressure of 1 bar; red: same as blue, but for 10 bar surface partial pressure of CO$_2$. Bottom: atmospheric models consistent with current observations of the Trappist 1b planet.  
   }
   \end{figure}

To further illustrate the points above we show in Figure~\ref{fig:7} (top panel) example atmospheres for habitable planets with different compositions seen from 10 pc distance. The atmospheric temperature profiles are self-consistently calculated using the atmospheric model described in Kitzmann (2016)\cite{Kitzmann2010A&A...511A..66K} and Kitzmann (2017)\cite{Kitzmann2017A&A...600A.111K}. For this study, the model has been expanded to include additional molecules, such as O$_3$, N$_2$O, and CH$_4$. For the low-temperature Earth-like cases, the HITRAN linelists are used to derive the corresponding molecular absorption cross sections. For the high-temperature cases of Trappist-1b (bottom panel of Figure~\ref{fig:7}), we use the Exomol linelists for all considered molecules. The cross sections for these very large linelists have been generated by the \textsc{Helios-k} opacity calculator\cite{Grimm2015ApJ...808..182G}.
The habitable, non-Earth-like scenarios assume a N$_2$ surface partial pressure of 1 bar, while the CO$_2$ surface partial pressure is varied between 1 and 10 bar. Both molecules are assumed to be well-mixed throughout the atmosphere. The atmospheric H$_2$O profile is calculated as a function of the surface temperature, assuming an Earth-like relative humidity profile in the troposphere\cite{Kitzmann2017A&A...600A.111K}. The orbital distances are chosen to obtain an Earth-like surface temperature of 288 K in each case. In addition, an Earth-twin atmosphere is included. The atmospheric chemical composition, including ozone and methane, is taken from Kitzmann et al. (2010)\cite{Kitzmann2010A&A...511A..66K}. 
In the bottom panel, which shows potential atmospheres for the Trappist 1b planet consistent with recent observations\cite{delrez2018}, all considered atmospheric constituents are assumed to have isoprofiles throughout the entire atmosphere. The stellar insulation for Trappist 1b is roughly 4 times the solar constant resulting in an estimated equilibrium temperature of $\sim$400 K, while the surface temperature can reach values well above 1000 K for the water-rich cases. Not only is its overall flux expected to be significantly higher than that of the planets in the panel above, but also its emission peak is at shorter wavelength irrespective of the assumed composition. 

\subsection{Spectral resolution requirements}
Determining the optimal spectral resolution of an MIR nulling interferometer is more challenging and should be done in the context of a comprehensive retrieval study. As most features are broad the question is more about spectral contamination from neighboring features. As mentioned above, N$_2$O and CH$_4$ are in principle detectable at MIR wavelengths, but they are both located in the 7--9\,$\mu$m range\footnote{N$_2$O has another weaker feature at $\sim$17\,$\mu$m.} next to a potentially dominant water feature\cite{desmarais2002}. It will be important to quantify and re-assess what levels of N$_2$O and/or CH$_4$ can be detected as a function of spectral resolution and water content. The same is true for the collision-induced (N$_2$)$_2$ feature in the wing of the $\sim$4.3\,$\mu$m CO$_2$band\cite{schwieterman2015}.

\section{TECHNOLOGY READINESS AND DEVELOPMENT}
%- where does key technology for space-based MIR interferometry stand today compared to Darwin/TPF-I times \\
%- what will happen in the near term (i.e., what is already planned\\
%- what are key things that still need to be done (e.g., null depth + stability under cryo conditions, integrated optics in MIR).

Where do we stand in terms of technology? Most technologies required to fly a space-based nulling interferometer have today reached a Technology Readiness Level (TRL) of 5, which means that the components have been tested and validated in relevant environment. In particular, key technologies that were considered immature in 2007, when most Darwin/TPF-I activities stopped, have now been demonstrated on test-benches (e.g., deep nulling beam combination) or in space (e.g., formation flying). Below we review briefly the major progress achieved since 2007 as well as current developments with ground-based interferometric facilities. A recent review on technology status can also be found in Defr\`ere et al. 2017\cite{defrere2017}.

\subsection{Formation flying}

Remarkable advances in technology have been made in Europe in recent years with the space-based demonstration of this technology by the PRISMA mission (\url{http://www.snsb.se/en/Home/Space-Activities-in-Sweden/Satellites/}). PRISMA demonstrated a sub-cm positioning accuracy between two spacecraft, mainly limited by the metrology system (GPS and RF). The launch of ESA's PROBA-3 mission in 2019 will provide further valuable free-flyer positioning accuracy results (sub-mm), which exceeds the requirements for a space-based nulling interferometer. Extending the flight-tested building-block functionality from a distributed two-spacecraft instrument to an instrument with more spacecraft mainly relies on the replication of the coordination functionality and does not present additional complexity in terms of procedures according to the PRISMA navigation team. While formation flying can then be considered to have reached TRL 9, once PROBA-3 has flown, an uncertainty remains regarding fuel usage and the possible lifetime of such a mission. 

\subsection{Starlight suppression}

A considerable expertise has been developed on starlight suppression over the past 20 years, both in academic and industrial centers across the globe. Approximately 35 PhD theses were dedicated to this topic and more than 40 refereed papers. These efforts culminated with laboratory demonstrations at room temperature mainly at the \emph{Jet Propulsion Laboratory} (JPL) in the US. For instance, work  with  the  Adaptive  Nuller  has  indicated  that  mid-infrared nulls of 10$^{-5}$ are achievable with a bandwidth of 34\% and a mean wavelength of 10\,$\mu$m \cite{Peters:2010}. Another testbed, the planet detection testbed, was developed in parallel and demonstrated the  main components of a high performance four-beam nulling interferometer at a level matching that needed for the space mission \cite{Martin:2010}. At 10~$\mu$m with 10\% bandwidth, it has achieved nulling of $8\times 10^{-6}$ (the flight requirement is $10^{-5}$), starlight suppression of $10^{-8}$ after post-processing, and actual planet detection at a planet-to-star contrast of $3\times 10^{-7}$, i.e., the Earth-Sun contrast. The phase chopping technique \cite{Mennesson:2005} has also been implemented and validated on-sky with the Keck Nuller Interferometer \cite{Colavita:2009}. A null stability of a few $\sim$10$^{-3}$ was achieved, mainly limited by the large thermal background and variable water vapor content, both effects specific to ground-based mid-infrared observations.

\subsection{Current developments}

Over the past decade, the operation of high-precision ground-based interferometers has matured in both Europe and the US. Europe has gained a strong expertise in the field of fringe sensing, tracking, and stabilization with the operation of the Very Large Telescope Interferometer (VLTI). In the United States, a lot of technical expertise was gained by operating several nulling interferometers such as the Keck Interferometer Nuller \cite{Colavita:2009}, the Palomar Fiber Nuller \cite{Mennesson:2011a}, and the Large Binocular Telescope Interferometer \cite{Hinz:2014}. All have produced excellent scientific results \cite{Mennesson:2014,Defrere:2015,Ertel:2018} and pushed high-resolution mid-infrared imaging to new limits \cite{Defrere:2016}. New innovative data reduction techniques have also been developed to improve the accuracy of nulling instruments \cite{Hanot:2011,Mennesson:2011b} but more work is required to adapt this technique to four-telescope configurations.

Interferometric telescope arrays operating at infrared wavelengths also provide opportunities for testing technologies on the ground and to push new developments ahead in collaboration with the existing interferometry community. Within this community, there is currently a science-driven international initiative to develop a new high-contrast interferometric instrument for the VLTI \cite{Defrere:2018} to detect young exoplanets and measure the prevalence of exozodiacal dust around Southern stars (not covered by the LBTI). This project will serve as a natural scientific and technology precursor for a large space-based mission. There is also an ongoing collaboration (The Planet Formation Imager -- PFI\footnote{http://planetformationimager.org}) to establish a road-map for a future ground-based facility that will be optimized to image planet-forming disks on the spatial scales where protoplanets are assembled (i.e., the Hill sphere of the forming planets). With $\sim 20$ telescope elements and baselines of $\sim 3$~km, the PFI concept is optimized for imaging complex scenes at mid-infrared wavelengths ($3-12\mu$m) and at 0.1 milliarcsecond resolution, complementing the capabilities of a space interferometer that would be optimized to achieve the sensitivity and contrast required to characterize the atmospheres of mature exoplanets. 

\subsection{Required technology developments}

Starlight suppression has already been demonstrated to flight requirements \cite{Martin:2010}, but with fluxes much higher than those expected from stars and planets allowing working at room temperature without being disturbed by the thermal emission of the environment. The next step would be to reproduce this experiment in cryogenic conditions and with optical fluxes similar to the astronomical ones. This will require in particular the successful validation of cryogenic spatial filters that can provide the necessary performance from 6 (or 3) to 20~$\mu$m and the implementation of a cryogenic deformable mirror, which is now within reach \cite{Enya:2009}. The spatial filtering capabilities of photonic crystal fibers should also be investigated for use at MIR wavelengths, because of the improved throughput that they may provide and the possibility to cover the whole wavelength range with a single technology. Note finally that, with recent developments in wavefront control with extreme adaptive optics systems \cite{Jovanovic:2015}, it is not clear whether such a technology will be required. This should be addressed in the future. Specific developments in terms of fringe tracking (taking into account residual vibrations) and data reduction will undoubtedly be needed to reach the required level of performance in terms of starlight rejection. We also expect that dedicated developments will be required in the field of MIR detectors, although the JWST legacy will be particularly useful in this context.

\section{CONCLUSIONS AND NEXT STEPS}
The Monte Carlo simulations shown here demonstrate that a MIR space-based nulling interferometer, as once studied in the context of ESA's Darwin and NASA's TPF-I missions, could yield at least as many exoplanet detections as a large, single aperture optical/NIR telescope. The details and exact number of planets certainly depends on the assumed technical specifications and the underlying exoplanet populations, but from an exoplanet science perspective such an interferometer should be considered an  attractive mission concept, at least complementary if not superior to an optical/NIR mission. 

Our analysis also shows that getting a better handle on the overall planet statistics is crucial for planning larger future missions. The significant differences in expected planet yield for Population 1 and 2 need to be investigated and -- if possible -- rectified. 

Another key aspect that we will investigate more closely in the future is a specific treatment of stellar leakage and exozodical light in our simulations. One way to do that is to couple our Monte Carlo simulations with {\tt DarwinSIM}\cite{Defrere:2010}, a dedicated simulator for space-based nulling interferometry, and include our latest understanding of the occurrence rate and levels of exozodi emission. The usage of {\tt DarwinSIM} would further allow us to study different designs and formations for the interferometer (e.g., 4 vs. 3 collector telescopes) and the impact on the science yield, and a first optimization concerning the appropriate time-per-target would also be possible. A critical look at the stellar input sample and its properties is also warranted with a specific focus on multiplicity. Finally, a more detailed investigation concerning the science requirements for the spectral resolution and the wavelength rage needs to be carried out and one should re-assess what band-passes should be used for the ``discovery phase'' of the mission. A dedicated and comprehensive spectral retrieval study would be one way to address these points. 

On the technical side it is worth noting that significant progress has been made for some of the major challenges even after Darwin and TPF-I were canceled. Now it is about time to consolidate these findings, identify the remaining missing pieces and craft a road-map for further technology development that will -- eventually -- lead to a mission that can successfully address the most fundamental questions of exoplanet science. 

\acknowledgments % equivalent to \section*{ACKNOWLEDGMENTS}       
Parts of this work have been carried out within the framework of the National Center for Competence in Research PlanetS supported by the SNSF. SPQ acknowledges the financial support of the SNSF. The authors would like to thank A. Zsom for valuable support during the analysis using the {\tt HUNTER} code and T. Boulet for providing information about the detectability of known exoplanets. 
% References
\bibliographystyle{spiebib} % makes bibtex use spiebib.bst

\begin{thebibliography}{10}

\bibitem{mayorqueloz1995}
{Mayor}, M. and {Queloz}, D., ``{A Jupiter-mass companion to a solar-type
  star},'' {\em Nature}~{\bf 378},  355--359 (Nov. 1995).

\bibitem{burke2015}
{Burke}, C.~J., {Christiansen}, J.~L., {Mullally}, F., {Seader}, S., {Huber},
  D., {Rowe}, J.~F., {Coughlin}, J.~L., {Thompson}, S.~E., {Catanzarite}, J.,
  {Clarke}, B.~D., {Morton}, T.~D., {Caldwell}, D.~A., {Bryson}, S.~T., {Haas},
  M.~R., {Batalha}, N.~M., {Jenkins}, J.~M., {Tenenbaum}, P., {Twicken}, J.~D.,
  {Li}, J., {Quintana}, E., {Barclay}, T., {Henze}, C.~E., {Borucki}, W.~J.,
  {Howell}, S.~B., and {Still}, M., ``{Terrestrial Planet Occurrence Rates for
  the Kepler GK Dwarf Sample},'' {\em ApJ}~{\bf 809},  8 (Aug. 2015).

\bibitem{mayor2011}
{Mayor}, M., {Marmier}, M., {Lovis}, C., {Udry}, S., {S{\'e}gransan}, D.,
  {Pepe}, F., {Benz}, W., {Bertaux}, J.~., {Bouchy}, F., {Dumusque}, X., {Lo
  Curto}, G., {Mordasini}, C., {Queloz}, D., and {Santos}, N.~C., ``{The HARPS
  search for southern extra-solar planets XXXIV. Occurrence, mass distribution
  and orbital properties of super-Earths and Neptune-mass planets},'' {\em
  ArXiv e-prints}  (Sept. 2011).

\bibitem{bonfils2013}
{Bonfils}, X., {Delfosse}, X., {Udry}, S., {Forveille}, T., {Mayor}, M.,
  {Perrier}, C., {Bouchy}, F., {Gillon}, M., {Lovis}, C., {Pepe}, F., {Queloz},
  D., {Santos}, N.~C., {S{\'e}gransan}, D., and {Bertaux}, J.-L., ``{The HARPS
  search for southern extra-solar planets. XXXI. The M-dwarf sample},'' {\em
  A\&A}~{\bf 549},  A109 (Jan. 2013).

\bibitem{seager2010}
{Seager}, S. and {Deming}, D., ``{Exoplanet Atmospheres},'' {\em ARAA}~{\bf
  48},  631--672 (Sept. 2010).

\bibitem{sing2016}
{Sing}, D.~K., {Fortney}, J.~J., {Nikolov}, N., {Wakeford}, H.~R., {Kataria},
  T., {Evans}, T.~M., {Aigrain}, S., {Ballester}, G.~E., {Burrows}, A.~S.,
  {Deming}, D., {D{\'e}sert}, J.-M., {Gibson}, N.~P., {Henry}, G.~W.,
  {Huitson}, C.~M., {Knutson}, H.~A., {Lecavelier Des Etangs}, A., {Pont}, F.,
  {Showman}, A.~P., {Vidal-Madjar}, A., {Williamson}, M.~H., and {Wilson},
  P.~A., ``{A continuum from clear to cloudy hot-Jupiter exoplanets without
  primordial water depletion},'' {\em Nature}~{\bf 529},  59--62 (Jan. 2016).

\bibitem{kreidberg2014}
{Kreidberg}, L., {Bean}, J.~L., {D{\'e}sert}, J.-M., {Benneke}, B., {Deming},
  D., {Stevenson}, K.~B., {Seager}, S., {Berta-Thompson}, Z., {Seifahrt}, A.,
  and {Homeier}, D., ``{Clouds in the atmosphere of the super-Earth exoplanet
  GJ1214b},'' {\em Nature}~{\bf 505},  69--72 (Jan. 2014).

\bibitem{dewit2016}
{de Wit}, J., {Wakeford}, H.~R., {Gillon}, M., {Lewis}, N.~K., {Valenti},
  J.~A., {Demory}, B.-O., {Burgasser}, A.~J., {Burdanov}, A., {Delrez}, L.,
  {Jehin}, E., {Lederer}, S.~M., {Queloz}, D., {Triaud}, A.~H.~M.~J., and {Van
  Grootel}, V., ``{A combined transmission spectrum of the Earth-sized
  exoplanets TRAPPIST-1 b and c},'' {\em \nat}~{\bf 537},  69--72 (Sept. 2016).

\bibitem{morley2017}
{Morley}, C.~V., {Kreidberg}, L., {Rustamkulov}, Z., {Robinson}, T., and
  {Fortney}, J.~J., ``{Observing the Atmospheres of Known Temperate Earth-sized
  Planets with JWST},'' {\em ApJ}~{\bf 850},  121 (Dec. 2017).

\bibitem{sullivan2015}
{Sullivan}, P.~W., {Winn}, J.~N., {Berta-Thompson}, Z.~K., {Charbonneau}, D.,
  {Deming}, D., {Dressing}, C.~D., {Latham}, D.~W., {Levine}, A.~M.,
  {McCullough}, P.~R., {Morton}, T., {Ricker}, G.~R., {Vanderspek}, R., and
  {Woods}, D., ``{The Transiting Exoplanet Survey Satellite: Simulations of
  Planet Detections and Astrophysical False Positives},'' {\em ApJ}~{\bf 809},
  77 (Aug. 2015).

\bibitem{cessa2017}
{Cessa}, V., {Beck}, T., {Benz}, W., {Broeg}, C., {Ehrenreich}, D., {Fortier},
  A., {Peter}, G., {Magrin}, D., {Pagano}, I., {Plesseria}, J.-Y., {Steller},
  M., {Szoke}, J., {Thomas}, N., {Ragazzoni}, R., and {Wildi}, F., ``{CHEOPS: a
  space telescope for ultra-high precision photometry of exoplanet transits},''
  in [{\em Society of Photo-Optical Instrumentation Engineers (SPIE) Conference
  Series}{\nolinebreak\hspace{0.1em}]},  {\em Society of Photo-Optical
  Instrumentation Engineers (SPIE) Conference Series} {\bf 10563},  105631L
  (Nov. 2017).

\bibitem{rauer2014}
{Rauer}, H., {Catala}, C., {Aerts}, C., {Appourchaux}, T., {Benz}, W.,
  {Brandeker}, A., {Christensen-Dalsgaard}, J., {Deleuil}, M., {Gizon}, L.,
  {Goupil}, M.-J., {G{\"u}del}, M., {Janot-Pacheco}, E., {Mas-Hesse}, M.,
  {Pagano}, I., {Piotto}, G., {Pollacco}, D., {Santos}, {\.C}., {Smith}, A.,
  {Su{\'a}rez}, J.-C., {Szab{\'o}}, R., {Udry}, S., {Adibekyan}, V., {Alibert},
  Y., {Almenara}, J.-M., {Amaro-Seoane}, P., {Eiff}, M.~A.-v., {Asplund}, M.,
  {Antonello}, E., {Barnes}, S., {Baudin}, F., {Belkacem}, K., {Bergemann}, M.,
  {Bihain}, G., {Birch}, A.~C., {Bonfils}, X., {Boisse}, I., {Bonomo}, A.~S.,
  {Borsa}, F., {Brand{\~a}o}, I.~M., {Brocato}, E., {Brun}, S., {Burleigh}, M.,
  {Burston}, R., {Cabrera}, J., {Cassisi}, S., {Chaplin}, W., {Charpinet}, S.,
  {Chiappini}, C., {Church}, R.~P., {Csizmadia}, S., {Cunha}, M., {Damasso},
  M., {Davies}, M.~B., {Deeg}, H.~J., {D{\'{\i}}az}, R.~F., {Dreizler}, S.,
  {Dreyer}, C., {Eggenberger}, P., {Ehrenreich}, D., {Eigm{\"u}ller}, P.,
  {Erikson}, A., {Farmer}, R., {Feltzing}, S., {de Oliveira Fialho}, F.,
  {Figueira}, P., {Forveille}, T., {Fridlund}, M., {Garc{\'{\i}}a}, R.~A.,
  {Giommi}, P., {Giuffrida}, G., {Godolt}, M., {Gomes da Silva}, J., {Granzer},
  T., {Grenfell}, J.~L., {Grotsch-Noels}, A., {G{\"u}nther}, E., {Haswell},
  C.~A., {Hatzes}, A.~P., {H{\'e}brard}, G., {Hekker}, S., {Helled}, R.,
  {Heng}, K., {Jenkins}, J.~M., {Johansen}, A., {Khodachenko}, M.~L.,
  {Kislyakova}, K.~G., {Kley}, W., {Kolb}, U., {Krivova}, N., {Kupka}, F.,
  {Lammer}, H., {Lanza}, A.~F., {Lebreton}, Y., {Magrin}, D., {Marcos-Arenal},
  P., {Marrese}, P.~M., {Marques}, J.~P., {Martins}, J., {Mathis}, S.,
  {Mathur}, S., {Messina}, S., {Miglio}, A., {Montalban}, J., {Montalto}, M.,
  {Monteiro}, M.~J.~P.~F.~G., {Moradi}, H., {Moravveji}, E., {Mordasini}, C.,
  {Morel}, T., {Mortier}, A., {Nascimbeni}, V., {Nelson}, R.~P., {Nielsen},
  M.~B., {Noack}, L., {Norton}, A.~J., {Ofir}, A., {Oshagh}, M., {Ouazzani},
  R.-M., {P{\'a}pics}, P., {Parro}, V.~C., {Petit}, P., {Plez}, B., {Poretti},
  E., {Quirrenbach}, A., {Ragazzoni}, R., {Raimondo}, G., {Rainer}, M.,
  {Reese}, D.~R., {Redmer}, R., {Reffert}, S., {Rojas-Ayala}, B., {Roxburgh},
  I.~W., {Salmon}, S., {Santerne}, A., {Schneider}, J., {Schou}, J., {Schuh},
  S., {Schunker}, H., {Silva-Valio}, A., {Silvotti}, R., {Skillen}, I.,
  {Snellen}, I., {Sohl}, F., {Sousa}, S.~G., {Sozzetti}, A., {Stello}, D.,
  {Strassmeier}, K.~G., {{\v S}vanda}, M., {Szab{\'o}}, G.~M., {Tkachenko}, A.,
  {Valencia}, D., {Van Grootel}, V., {Vauclair}, S.~D., {Ventura}, P.,
  {Wagner}, F.~W., {Walton}, N.~A., {Weingrill}, J., {Werner}, S.~C.,
  {Wheatley}, P.~J., and {Zwintz}, K., ``{The PLATO 2.0 mission},'' {\em
  Experimental Astronomy}~{\bf 38},  249--330 (Nov. 2014).

\bibitem{tinetti2016}
{Tinetti}, G., {Drossart}, P., {Eccleston}, P., {Hartogh}, P., {Heske}, A.,
  {Leconte}, J., {Micela}, G., {Ollivier}, M., {Pilbratt}, G., {Puig}, L.,
  {Turrini}, D., {Vandenbussche}, B., {Wolkenberg}, P., {Pascale}, E.,
  {Beaulieu}, J.-P., {G{\"u}del}, M., {Min}, M., {Rataj}, M., {Ray}, T.,
  {Ribas}, I., {Barstow}, J., {Bowles}, N., {Coustenis}, A., {Coud{\'e} du
  Foresto}, V., {Decin}, L., {Encrenaz}, T., {Forget}, F., {Friswell}, M.,
  {Griffin}, M., {Lagage}, P.~O., {Malaguti}, P., {Moneti}, A., {Morales},
  J.~C., {Pace}, E., {Rocchetto}, M., {Sarkar}, S., {Selsis}, F., {Taylor}, W.,
  {Tennyson}, J., {Venot}, O., {Waldmann}, I.~P., {Wright}, G., {Zingales}, T.,
  and {Zapatero-Osorio}, M.~R., ``{The science of ARIEL (Atmospheric
  Remote-sensing Infrared Exoplanet Large-survey)},'' in [{\em Space Telescopes
  and Instrumentation 2016: Optical, Infrared, and Millimeter
  Wave}{\nolinebreak\hspace{0.1em}]},  {\em SPIE} {\bf 9904},  99041X (July
  2016).

\bibitem{noecker2016}
{Noecker}, M.~C., {Zhao}, F., {Demers}, R., {Trauger}, J., {Guyon}, O., and
  {Jeremy Kasdin}, N., ``{Coronagraph instrument for WFIRST-AFTA},'' {\em
  Journal of Astronomical Telescopes, Instruments, and Systems}~{\bf 2},
  011001 (Jan. 2016).

\bibitem{brandl2016}
{Brandl}, B.~R., {Ag{\'o}cs}, T., {Aitink-Kroes}, G., {Bertram}, T.,
  {Bettonvil}, F., {van Boekel}, R., {Boulade}, O., {Feldt}, M., {Glasse}, A.,
  {Glauser}, A., {G{\"u}del}, M., {Hurtado}, N., {Jager}, R., {Kenworthy},
  M.~A., {Mach}, M., {Meisner}, J., {Meyer}, M., {Pantin}, E., {Quanz}, S.,
  {Schmid}, H.~M., {Stuik}, R., {Veninga}, A., and {Waelkens}, C., ``{Status of
  the mid-infrared E-ELT imager and spectrograph METIS},'' in [{\em
  Ground-based and Airborne Instrumentation for Astronomy
  VI}{\nolinebreak\hspace{0.1em}]},  {\em SPIE} {\bf 9908},  990820 (Aug.
  2016).

\bibitem{marconi2016}
{Marconi}, A., {Di Marcantonio}, P., {D'Odorico}, V., {Cristiani}, S.,
  {Maiolino}, R., {Oliva}, E., {Origlia}, L., {Riva}, M., {Valenziano}, L.,
  {Zerbi}, F.~M., {Abreu}, M., {Adibekyan}, V., {Allende Prieto}, C., {Amado},
  P.~J., {Benz}, W., {Boisse}, I., {Bonfils}, X., {Bouchy}, F., {Buchhave}, L.,
  {Buscher}, D., {Cabral}, A., {Canto Martins}, B.~L., {Chiavassa}, A.,
  {Coelho}, J., {Christensen}, L.~B., {Delgado-Mena}, E., {de Medeiros}, J.~R.,
  {Di Varano}, I., {Figueira}, P., {Fisher}, M., {Fynbo}, J.~P.~U., {Glasse},
  A.~C.~H., {Haehnelt}, M., {Haniff}, C., {Hansen}, C.~J., {Hatzes}, A.,
  {Huke}, P., {Korn}, A.~J., {Le{\~a}o}, I.~C., {Liske}, J., {Lovis}, C.,
  {Maslowski}, P., {Matute}, I., {McCracken}, R.~A., {Martins}, C.~J.~A.~P.,
  {Monteiro}, M.~J.~P.~F.~G., {Morris}, S., {Morris}, T., {Nicklas}, H.,
  {Niedzielski}, A., {Nunes}, N.~J., {Palle}, E., {Parr-Burman}, P.~M.,
  {Parro}, V., {Parry}, I., {Pepe}, F., {Piskunov}, N., {Queloz}, D.,
  {Quirrenbach}, A., {Rebolo Lopez}, R., {Reiners}, A., {Reid}, D.~T.,
  {Santos}, N., {Seifert}, W., {Sousa}, S., {Stempels}, H.~C., {Strassmeier},
  K., {Sun}, X., {Udry}, S., {Vanzi}, L., {Vestergaard}, M., {Weber}, M., and
  {Zackrisson}, E., ``{EELT-HIRES the high-resolution spectrograph for the
  E-ELT},'' in [{\em Ground-based and Airborne Instrumentation for Astronomy
  VI}{\nolinebreak\hspace{0.1em}]},  {\em SPIE} {\bf 9908},  990823 (Aug.
  2016).

\bibitem{verniaud2010}
{V{\'e}rinaud}, C., {Kasper}, M., {Beuzit}, J.-L., {Gratton}, R.~G., {Mesa},
  D., {Aller-Carpentier}, E., {Fedrigo}, E., {Abe}, L., {Baudoz}, P.,
  {Boccaletti}, A., {Bonavita}, M., {Dohlen}, K., {Hubin}, N., {Kerber}, F.,
  {Korkiakoski}, V., {Antichi}, J., {Martinez}, P., {Rabou}, P., {Roelfsema},
  R., {Schmid}, H.~M., {Thatte}, N., {Salter}, G., {Tecza}, M., {Venema}, L.,
  {Hanenburg}, H., {Jager}, R., {Yaitskova}, N., {Preis}, O., {Orecchia}, M.,
  and {Stadler}, E., ``{System study of EPICS: the exoplanets imager for the
  E-ELT},'' in [{\em Adaptive Optics Systems II}{\nolinebreak\hspace{0.1em}]},
  {\em SPIE} {\bf 7736},  77361N (July 2010).

\bibitem{quanz2015}
{Quanz}, S.~P., {Crossfield}, I., {Meyer}, M.~R., {Schmalzl}, E., and {Held},
  J., ``{Direct detection of exoplanets in the 3-10 {$\mu$}m range with
  E-ELT/METIS},'' {\em International Journal of Astrobiology}~{\bf 14},
  279--289 (Apr. 2015).

\bibitem{gonzalezhernandez2017}
{Gonz{\'a}lez Hern{\'a}ndez}, J.~I., {Pepe}, F., {Molaro}, P., and {Santos},
  N., ``{ESPRESSO on VLT: An Instrument for Exoplanet Research},'' {\em ArXiv
  e-prints}  (Nov. 2017).

\bibitem{messesson2016}
{Mennesson}, B., {Gaudi}, S., {Seager}, S., {Cahoy}, K., {Domagal-Goldman}, S.,
  {Feinberg}, L., {Guyon}, O., {Kasdin}, J., {Marois}, C., {Mawet}, D.,
  {Tamura}, M., {Mouillet}, D., {Prusti}, T., {Quirrenbach}, A., {Robinson},
  T., {Rogers}, L., {Scowen}, P., {Somerville}, R., {Stapelfeldt}, K., {Stern},
  D., {Still}, M., {Turnbull}, M., {Booth}, J., {Kiessling}, A., {Kuan}, G.,
  and {Warfield}, K., ``{The Habitable Exoplanet (HabEx) Imaging Mission:
  preliminary science drivers and technical requirements},'' in [{\em Space
  Telescopes and Instrumentation 2016: Optical, Infrared, and Millimeter
  Wave}{\nolinebreak\hspace{0.1em}]},  {\em SPIE} {\bf 9904},  99040L (July
  2016).

\bibitem{peterson2017}
{Peterson}, B.~M., {Fischer}, D., and {LUVOIR Science and Technology Definition
  Team}, ``{The Large Ultraviolet/Optical/Infrared Surveyor (LUVOIR)},'' in
  [{\em American Astronomical Society Meeting Abstracts
  \#229}{\nolinebreak\hspace{0.1em}]},  {\em American Astronomical Society
  Meeting Abstracts} {\bf 229},  405.04 (Jan. 2017).

\bibitem{kammerer2018}
{Kammerer}, J. and {Quanz}, S.~P., ``{Simulating the exoplanet yield of a
  space-based mid-infrared interferometer based on Kepler statistics},'' {\em
  A\&A}~{\bf 609},  A4 (Jan. 2018).

\bibitem{Hsu2018}
{Hsu}, D.~C., {Ford}, E.~B., {Ragozzine}, D., and {Morehead}, R.~C.,
  ``{Improving the Accuracy of Planet Occurrence Rates from Kepler Using
  Approximate Bayesian Computation},'' {\em \aj}~{\bf 155},  205 (May 2018).

\bibitem{cockell2009}
{Cockell}, C.~S., {Herbst}, T., {L{\'e}ger}, A., {Absil}, O., {Beichman}, C.,
  {Benz}, W., {Brack}, A., {Chazelas}, B., {Chelli}, A., {Cottin}, H.,
  {Coud{\'e} du Foresto}, V., {Danchi}, W., {Defr{\`e}re}, D., {den Herder},
  J.-W., {Eiroa}, C., {Fridlund}, M., {Henning}, T., {Johnston}, K.,
  {Kaltenegger}, L., {Labadie}, L., {Lammer}, H., {Launhardt}, R., {Lawson},
  P., {Lay}, O.~P., {Liseau}, R., {Martin}, S.~R., {Mawet}, D., {Mourard}, D.,
  {Moutou}, C., {Mugnier}, L., {Paresce}, F., {Quirrenbach}, A., {Rabbia}, Y.,
  {Rottgering}, H.~J.~A., {Rouan}, D., {Santos}, N., {Selsis}, F., {Serabyn},
  E., {Westall}, F., {White}, G., {Ollivier}, M., and {Bord{\'e}}, P.,
  ``{Darwin -- an experimental astronomy mission to search for extrasolar
  planets},'' {\em Experimental Astronomy}~{\bf 23},  435--461 (Mar. 2009).

\bibitem{glasse2015}
{Glasse}, A., {Rieke}, G.~H., {Bauwens}, E., {Garc{\'{\i}}a-Mar{\'{\i}}n}, M.,
  {Ressler}, M.~E., {Rost}, S., {Tikkanen}, T.~V., {Vandenbussche}, B., and
  {Wright}, G.~S., ``{The Mid-Infrared Instrument for the James Webb Space
  Telescope, IX: Predicted Sensitivity},'' {\em \pasp}~{\bf 127},  686 (July
  2015).

\bibitem{defrere2010}
{Defr{\`e}re}, D., {Absil}, O., {den Hartog}, R., {Hanot}, C., and {Stark}, C.,
  ``{Nulling interferometry: impact of exozodiacal clouds on the performance of
  future life-finding space missions},'' {\em \aap}~{\bf 509},  A9 (Jan. 2010).

\bibitem{defrere2012}
{Defr{\`e}re}, D., {Stark}, C., {Cahoy}, K., and {Beerer}, I., ``{Direct
  imaging of exoEarths embedded in clumpy debris disks},'' in [{\em Space
  Telescopes and Instrumentation 2012: Optical, Infrared, and Millimeter
  Wave}{\nolinebreak\hspace{0.1em}]},  {\em \procspie} {\bf 8442},  84420M
  (Sept. 2012).

\bibitem{rogers2015}
Rogers, L.~A., ``{MOST 1.6 EARTH-RADIUS PLANETS ARE NOT ROCKY},'' {\em The
  Astrophysical Journal}~{\bf 801},  41 (Mar. 2015).

\bibitem{chen2017}
Chen, J. and Kipping, D., ``{Probabilistic Forecasting of the Masses and Radii
  of Other Worlds},'' {\em The Astrophysical Journal}~{\bf 834},  17 (Jan.
  2017).

\bibitem{kaltenegger2017}
{Kaltenegger}, L., ``{How to Characterize Habitable Worlds and Signs of
  Life},'' {\em \araa}~{\bf 55},  433--485 (Aug. 2017).

\bibitem{seager2013}
{Seager}, S., {Bains}, W., and {Hu}, R., ``{Biosignature Gases in
  H$_{2}$-dominated Atmospheres on Rocky Exoplanets},'' {\em \apj}~{\bf 777},
  95 (Nov. 2013).

\bibitem{zsom2015}
{Zsom}, A., ``{A Population-based Habitable Zone Perspective},'' {\em
  \apj}~{\bf 813},  9 (Nov. 2015).

\bibitem{lay2007}
{Lay}, O.~P., {Martin}, S.~R., and {Hunyadi}, S.~L., ``{Planet-finding
  performance of the TPF-I Emma architecture},'' in [{\em Techniques and
  Instrumentation for Detection of Exoplanets
  III}{\nolinebreak\hspace{0.1em}]},  {\em \procspie} {\bf 6693},  66930A
  (Sept. 2007).

\bibitem{stark2015}
{Stark}, C.~C., {Roberge}, A., {Mandell}, A., {Clampin}, M., {Domagal-Goldman},
  S.~D., {McElwain}, M.~W., and {Stapelfeldt}, K.~R., ``{Lower Limits on
  Aperture Size for an ExoEarth Detecting Coronagraphic Mission},'' {\em
  \apj}~{\bf 808},  149 (Aug. 2015).

\bibitem{desmarais2002}
{Des Marais}, D.~J., {Harwit}, M.~O., {Jucks}, K.~W., {Kasting}, J.~F., {Lin},
  D.~N.~C., {Lunine}, J.~I., {Schneider}, J., {Seager}, S., {Traub}, W.~A., and
  {Woolf}, N.~J., ``{Remote Sensing of Planetary Properties and Biosignatures
  on Extrasolar Terrestrial Planets},'' {\em Astrobiology}~{\bf 2},  153--181
  (June 2002).

\bibitem{vonparis2013}
{von Paris}, P., {Hedelt}, P., {Selsis}, F., {Schreier}, F., and {Trautmann},
  T., ``{Characterization of potentially habitable planets: Retrieval of
  atmospheric and planetary properties from emission spectra},'' {\em
  \aap}~{\bf 551},  A120 (Mar. 2013).

\bibitem{feng2018}
{Feng}, Y.~K., {Robinson}, T.~D., {Fortney}, J.~J., {Lupu}, R.~E., {Marley},
  M.~S., {Lewis}, N.~K., {Macintosh}, B., and {Line}, M.~R., ``{Characterizing
  Earth Analogs in Reflected Light: Atmospheric Retrieval Studies for Future
  Space Telescopes},'' {\em \aj}~{\bf 155},  200 (May 2018).

\bibitem{schwieterman2015}
{Schwieterman}, E.~W., {Robinson}, T.~D., {Meadows}, V.~S., {Misra}, A., and
  {Domagal-Goldman}, S., ``{Detecting and Constraining N$_{2}$ Abundances in
  Planetary Atmospheres Using Collisional Pairs},'' {\em \apj}~{\bf 810},  57
  (Sept. 2015).

\bibitem{robinson2018}
{Robinson}, T.~D. and {Reinhard}, C.~T., ``{Earth as an Exoplanet},'' {\em
  ArXiv e-prints}  (Apr. 2018).

\bibitem{Kitzmann2010A&A...511A..66K}
{Kitzmann}, D., {Patzer}, A.~B.~C., {von Paris}, P., {Godolt}, M., {Stracke},
  B., {Gebauer}, S., {Grenfell}, J.~L., and {Rauer}, H., ``{Clouds in the
  atmospheres of extrasolar planets. I. Climatic effects of multi-layered
  clouds for Earth-like planets and implications for habitable zones},'' {\em
  \aap}~{\bf 511},  A66 (Feb. 2010).

\bibitem{Kitzmann2017A&A...600A.111K}
{Kitzmann}, D., ``{Clouds in the atmospheres of extrasolar planets. V. The
  impact of CO$_{2}$ ice clouds on the outer boundary of the habitable zone},''
  {\em \aap}~{\bf 600},  A111 (Apr. 2017).

\bibitem{Grimm2015ApJ...808..182G}
{Grimm}, S.~L. and {Heng}, K., ``{HELIOS-K: An Ultrafast, Open-source Opacity
  Calculator for Radiative Transfer},'' {\em \apj}~{\bf 808},  182 (Aug. 2015).

\bibitem{delrez2018}
{Delrez}, L., {Gillon}, M., {Triaud}, A.~H.~M.~J., {Demory}, B.-O., {de Wit},
  J., {Ingalls}, J.~G., {Agol}, E., {Bolmont}, E., {Burdanov}, A., {Burgasser},
  A.~J., {Carey}, S.~J., {Jehin}, E., {Leconte}, J., {Lederer}, S., {Queloz},
  D., {Selsis}, F., and {Van Grootel}, V., ``{Early 2017 observations of
  TRAPPIST-1 with Spitzer},'' {\em \mnras}~{\bf 475},  3577--3597 (Apr. 2018).

\bibitem{defrere2017}
{Defr{\`e}re}, D., {Absil}, O., and {Beichman}, C.,  [{\em {Interferometric
  Space Missions for Exoplanet Science: Legacy of
  Darwin/TPF}}{\nolinebreak\hspace{0.1em}]},  82 (2017).

\bibitem{Peters:2010}
{Peters}, R.~D., {Lay}, O.~P., and {Lawson}, P.~R., ``{Mid-Infrared Adaptive
  Nulling for the Detection of Earthlike Exoplanets},'' {\em PASP}~{\bf 122},
  85--92 (Jan. 2010).

\bibitem{Martin:2010}
{Martin}, S.~R. and {Booth}, A.~J., ``{Demonstration of exoplanet detection
  using an infrared telescope array},'' {\em A\&A}~{\bf 520},  A96 (Sept.
  2010).

\bibitem{Mennesson:2005}
{Mennesson}, B., {L{\'e}ger}, A., and {Ollivier}, M., ``{Direct detection and
  characterization of extrasolar planets: The Mariotti space interferometer},''
  {\em Icarus}~{\bf 178},  570--588 (Nov. 2005).

\bibitem{Colavita:2009}
{Colavita}, M.~M., {Serabyn}, E., {Millan-Gabet}, R., {Koresko}, C.~D.,
  {Akeson}, R.~L., {Booth}, A.~J., {Mennesson}, B.~P., {Ragland}, S.~D.,
  {Appleby}, E.~C., {Berkey}, B.~C., {Cooper}, A., {Crawford}, S.~L.,
  {Creech-Eakman}, M.~J., {Dahl}, W., {Felizardo}, C., {Garcia-Gathright},
  J.~I., {Gathright}, J.~T., {Herstein}, J.~S., {Hovland}, E.~E., {Hrynevych},
  M.~A., {Ligon}, E.~R., {Medeiros}, D.~W., {Moore}, J.~D., {Morrison}, D.,
  {Paine}, C.~G., {Palmer}, D.~L., {Panteleeva}, T., {Smith}, B., {Swain},
  M.~R., {Smythe}, R.~F., {Summers}, K.~R., {Tsubota}, K., {Tyau}, C.,
  {Vasisht}, G., {Wetherell}, E., {Wizinowich}, P.~L., and {Woillez}, J.~M.,
  ``{Keck Interferometer Nuller Data Reduction and On-Sky Performance},'' {\em
  PASP}~{\bf 121},  1120--1138 (Oct. 2009).

\bibitem{Mennesson:2011a}
{Mennesson}, B., {Serabyn}, E., {Hanot}, C., {Martin}, S.~R., {Liewer}, K., and
  {Mawet}, D., ``{New Constraints on Companions and Dust within a Few AU of
  Vega},'' {\em \apj}~{\bf 736},  14 (July 2011).

\bibitem{Hinz:2014}
{Hinz}, P., {Bailey}, V.~P., {Defr{\`e}re}, D., {Downey}, E., {Esposito}, S.,
  {Hill}, J., {Hoffmann}, W.~F., {Leisenring}, J., {Montoya}, M., {McMahon},
  T., {Puglisi}, A., {Skemer}, A., {Skrutskie}, M., {Vaitheeswaran}, V., and
  {Vaz}, A., ``{Commissioning the LBTI for use as a nulling interferometer and
  coherent imager},'' in [{\em Optical and Infrared Interferometry
  IV}{\nolinebreak\hspace{0.1em}]},  {\em Proc. SPIE} {\bf 9146},  91460T (July
  2014).

\bibitem{Mennesson:2014}
{Mennesson}, B., {Millan-Gabet}, R., {Serabyn}, E., {Colavita}, M.~M., {Absil},
  O., {Bryden}, G., {Wyatt}, M., {Danchi}, W., {Defr{\`e}re}, D., {Dor{\'e}},
  O., {Hinz}, P., {Kuchner}, M., {Ragland}, S., {Scott}, N., {Stapelfeldt}, K.,
  {Traub}, W., and {Woillez}, J., ``{Constraining the Exozodiacal Luminosity
  Function of Main-sequence Stars: Complete Results from the Keck Nuller
  Mid-infrared Surveys},'' {\em \apj}~{\bf 797},  119 (Dec. 2014).

\bibitem{Defrere:2015}
{Defr{\`e}re}, D., {Hinz}, P.~M., {Skemer}, A.~J., {Kennedy}, G.~M., {Bailey},
  V.~P., {Hoffmann}, W.~F., {Mennesson}, B., {Millan-Gabet}, R., {Danchi},
  W.~C., {Absil}, O., {Arbo}, P., {Beichman}, C., {Brusa}, G., {Bryden}, G.,
  {Downey}, E.~C., {Durney}, O., {Esposito}, S., {Gaspar}, A., {Grenz}, P.,
  {Haniff}, C., {Hill}, J.~M., {Lebreton}, J., {Leisenring}, J.~M., {Males},
  J.~R., {Marion}, L., {McMahon}, T.~J., {Montoya}, M., {Morzinski}, K.~M.,
  {Pinna}, E., {Puglisi}, A., {Rieke}, G., {Roberge}, A., {Serabyn}, E.,
  {Sosa}, R., {Stapeldfeldt}, K., {Su}, K., {Vaitheeswaran}, V., {Vaz}, A.,
  {Weinberger}, A.~J., and {Wyatt}, M.~C., ``{First-light LBT Nulling
  Interferometric Observations: Warm Exozodiacal Dust Resolved within a Few AU
  of {$\eta$} Crv},'' {\em \apj}~{\bf 799},  42 (Jan. 2015).

\bibitem{Ertel:2018}
{Ertel}, S., {Defr{\`e}re}, D., {Hinz}, P., {Mennesson}, B., {Kennedy}, G.~M.,
  {Danchi}, W.~C., {Gelino}, C., {Hill}, J.~M., {Hoffmann}, W.~F., {Rieke}, G.,
  {Shannon}, A., {Spalding}, E., {Stone}, J.~M., {Vaz}, A., {Weinberger},
  A.~J., {Willems}, P., {Absil}, O., {Arbo}, P., {Bailey}, V.~P., {Beichman},
  C., {Bryden}, G., {Downey}, E.~C., {Durney}, O., {Esposito}, S., {Gaspar},
  A., {Grenz}, P., {Haniff}, C.~A., {Leisenring}, J.~M., {Marion}, L.,
  {McMahon}, T.~J., {Millan-Gabet}, R., {Montoya}, M., {Morzinski}, K.~M.,
  {Pinna}, E., {Power}, J., {Puglisi}, A., {Roberge}, A., {Serabyn}, E.,
  {Skemer}, A.~J., {Stapelfeldt}, K., {Su}, K.~Y.~L., {Vaitheeswaran}, V., and
  {Wyatt}, M.~C., ``{The HOSTS Survey: Exozodiacal Dust Measurements for 30
  Stars},'' {\em \aj}~{\bf 155},  194 (May 2018).

\bibitem{Defrere:2016}
{Defr{\`e}re}, D., {Hinz}, P.~M., {Mennesson}, B., {Hoffmann}, W.~F.,
  {Millan-Gabet}, R., {Skemer}, A.~J., {Bailey}, V., {Danchi}, W.~C., {Downey},
  E.~C., {Durney}, O., {Grenz}, P., {Hill}, J.~M., {McMahon}, T.~J., {Montoya},
  M., {Spalding}, E., {Vaz}, A., {Absil}, O., {Arbo}, P., {Bailey}, H.,
  {Brusa}, G., {Bryden}, G., {Esposito}, S., {Gaspar}, A., {Haniff}, C.~A.,
  {Kennedy}, G.~M., {Leisenring}, J.~M., {Marion}, L., {Nowak}, M., {Pinna},
  E., {Powell}, K., {Puglisi}, A., {Rieke}, G., {Roberge}, A., {Serabyn}, E.,
  {Sosa}, R., {Stapeldfeldt}, K., {Su}, K., {Weinberger}, A.~J., and {Wyatt},
  M.~C., ``{Nulling Data Reduction and On-sky Performance of the Large
  Binocular Telescope Interferometer},'' {\em \apj}~{\bf 824},  66 (June 2016).

\bibitem{Hanot:2011}
{Hanot}, C., {Mennesson}, B., {Martin}, S., {Liewer}, K., {Loya}, F., {Mawet},
  D., {Riaud}, P., {Absil}, O., and {Serabyn}, E., ``{Improving Interferometric
  Null Depth Measurements using Statistical Distributions: Theory and First
  Results with the Palomar Fiber Nuller},'' {\em \apj}~{\bf 729},  110 (Mar.
  2011).

\bibitem{Mennesson:2011b}
{Mennesson}, B., {Hanot}, C., {Serabyn}, E., {Liewer}, K., {Martin}, S.~R., and
  {Mawet}, D., ``{High-contrast Stellar Observations within the Diffraction
  Limit at the Palomar Hale Telescope},'' {\em \apj}~{\bf 743},  178 (Dec.
  2011).

\bibitem{Defrere:2018}
{Defr{\`e}re}, D., {Absil}, O., {Berger}, J.-P., {Boulet}, T., {Danchi}, W.~C.,
  {Ertel}, S., {Gallenne}, A., {H{\'e}nault}, F., {Hinz}, P., {Huby}, E.,
  {Ireland}, M., {Kraus}, S., {Labadie}, L., {Le Bouquin}, J.-B., {Martin}, G.,
  {Matter}, A., {M{\'e}rand}, A., {Mennesson}, B., {Minardi}, S., {Monnier},
  J., {Norris}, B., {Orban de Xivry}, G., {Pedretti}, E., {Pott}, J.-U.,
  {Reggiani}, M., {Serabyn}, E., {Surdej}, J., {Tristram}, K.~R.~W., and
  {Woillez}, J., ``{The path towards high-contrast imaging with the VLTI: the
  Hi-5 project},'' {\em ArXiv e-prints}  (Jan. 2018).

\bibitem{Enya:2009}
{Enya}, K., {Kataza}, H., and {Bierden}, P., ``{A Micro Electrical Mechanical
  Systems (MEMS)-based Cryogenic Deformable Mirror},'' {\em \pasp}~{\bf 121},
  260--265 (Mar. 2009).

\bibitem{Jovanovic:2015}
{Jovanovic}, N., {Martinache}, F., {Guyon}, O., {Clergeon}, C., {Singh}, G.,
  {Kudo}, T., {Garrel}, V., {Newman}, K., {Doughty}, D., {Lozi}, J., {Males},
  J., {Minowa}, Y., {Hayano}, Y., {Takato}, N., {Morino}, J., {Kuhn}, J.,
  {Serabyn}, E., {Norris}, B., {Tuthill}, P., {Schworer}, G., {Stewart}, P.,
  {Close}, L., {Huby}, E., {Perrin}, G., {Lacour}, S., {Gauchet}, L.,
  {Vievard}, S., {Murakami}, N., {Oshiyama}, F., {Baba}, N., {Matsuo}, T.,
  {Nishikawa}, J., {Tamura}, M., {Lai}, O., {Marchis}, F., {Duchene}, G.,
  {Kotani}, T., and {Woillez}, J., ``{The Subaru Coronagraphic Extreme Adaptive
  Optics System: Enabling High-Contrast Imaging on Solar-System Scales},'' {\em
  \pasp}~{\bf 127},  890 (Sept. 2015).

\bibitem{Defrere:2010}
{Defr{\`e}re}, D., {Absil}, O., {den Hartog}, R., {Hanot}, C., and {Stark}, C.,
  ``{Nulling interferometry: impact of exozodiacal clouds on the performance of
  future life-finding space missions},'' {\em \aap}~{\bf 509},  A9 (Jan. 2010).

\end{thebibliography}

\end{document}